%% file: sensorfingerprint.tex
\documentclass[letterpaper,twocolumn,10pt]{article}
\pdfoutput=1
\usepackage{usenix,epsfig,endnotes}
\usepackage{graphicx,subfigure,array}
\usepackage{times,amsmath,amssymb}
\usepackage{footnote,verbatim,threeparttable}
\usepackage[usenames,dvipsnames]{color}
\usepackage[font=small,labelfont=bf]{caption}
\usepackage{multirow,algorithm,algorithmic}
\usepackage[hyphens]{url}
\usepackage[colorlinks,urlcolor=blue,breaklinks=true]{hyperref}
\usepackage{breakurl}
\usepackage{cite}
\usepackage{enumitem}

\newcommand{\paragraphb}[1]{\vspace{0.05in}\noindent{\bf #1} }
\DeclareMathOperator*{\argmin}{arg\,min}

\begin{document}

\date{}

\title{\Large \bf Exploring Ways To Mitigate Sensor-Based Smartphone Fingerprinting}

\author{
{\rm Anupam Das, Nikita Borisov,  Matthew Caesar}\\
University of Illinois at Urbana-Champaign
} 

\maketitle

\thispagestyle{empty}

\subsection*{Abstract}
Modern smartphones contain motion sensors, such as accelerometers and gyroscopes. These sensors have many useful applications; however, they can also 
be used to uniquely identify a phone by measuring anomalies in the signals, which are a result from manufacturing imperfections. Such measurements can 
be conducted surreptitiously in the browser and can be used to track users across applications, websites, and visits.

We analyze techniques to mitigate such device fingerprinting either by calibrating the sensors to eliminate the signal anomalies, or by adding noise 
that obfuscates the anomalies. To do this, we first develop a highly accurate fingerprinting mechanism that combines multiple motion sensors and makes 
use of (inaudible) audio stimulation to improve detection. We then collect measurements from a large collection of smartphones and evaluate the impact 
of calibration and obfuscation techniques on the classifier accuracy.


\input{intro}
\input{related}

\input{sensors}

\input{features}

\input{evaluation}

\input{sensitivity}
\input{countermeasure}
\input{limitation}

\input{conclusion}

{\footnotesize \bibliographystyle{acm}
\bibliography{bibliograph}}

\appendix
\input{appendix}

\end{document}

%% file: intro.tex
\section{Introduction}

Smartphones are equipped with motion sensors, such as accelerometers and gyroscopes, that are available to applications and website and enable a variety of novel uses. These same sensors, however, can threaten user privacy by enabling \emph{sensor fingerprinting}. Manufacturing imperfections result in each sensor having unique characteristics in their produced signal. These characteristics can be captured in the form of a fingerprint and used to track users across repeat visits. The sensor fingerprint can be used to supplement or replace other privacy-invasive tracking technologies, such as cookies, or canvas fingerprinting~\cite{canvas-fingerprint}. Since the fingerprint relies on the physical characteristics of a particular device, it is immune to defenses such as clearing cookies and private browsing modes. 

We carry out a detailed investigation the feasibility of fingerprinting of motion sensors in smartphones. Practical fingerprinting faces several challenges. 
During a typical web browsing session, a smart phone is either held in a user's hand, resulting in noisy motion inputs, or is resting on a flat surface, minimizing the amount of sensor input. Additionally, web APIs for accessing motion sensor data have significantly lower resolution than is available to the operating systems and applications. We show that, using machine learning techniques, it is possible to combine a large number of features from both the accelerometer and gyroscope sensor streams and produce highly accurate classification despite these challenges. In some cases, we can improve the classifier accuracy by using an inaudible sound, played through the speakers, to stimulate the motion sensors. We evaluate our techniques in a variety of lab settings; additionally, we collected data from volunteer participants over the web, capturing a wide variety of smartphone models and operating systems. In our experiments, a web browsing session lasting under a minute is still sufficient to generate a fingerprint that can be used in to recognize the phone in the future.  

We next investigate two potential countermeasures to sensor fingerprinting. First, we consider the use of \emph{calibration} to eliminate some of the error that results from manufacturing imperfections. Promisingly, we find that calibrating the accelerometer is easy and has a significant impact on classification accuracy. Gyroscope calibration, however, is more challenging without specialized equipment, and attempts to calibrate the gyroscope by hand do not result in an effective countermeasure.

An alternative countermeasure is \emph{obfuscation}, which introduces additional noise to the sensor readings in the hopes of hiding the natural errors. Obfuscation has the advantage of not requiring a calibration step; we find that by adding noise that is similar in magnitude to the natural errors that result from manufacturing, we can reduce the accuracy of fingerprinting more effectively than by calibration. We also investigate the possibility of using higher magnitude noise, as well as adding temporal disturbances to obfuscate frequency domain features. At high levels of noise, fingerprinting accuracy is greatly reduced, though such noise is likely to impair the utility of motion sensors.

\paragraphb{Roadmap.} The remainder of this paper is organized as follows. We present background information and related works in Section 
{\ref{background}}. In Section {\ref{sensors}}, we briefly discuss why accelerometers and gyroscopes can be used to generate 
unique fingerprints. In Section {\ref{feature_algo}}, we describe the different temporal and spectral features considered in our experiments, along 
with the classification algorithms and metrics used in our evaluations. We present our fingerprinting results in Section 
{\ref{evaluation}}. Section ~\ref{defense} describes our countermeasure techniques to sensor fingerprinting. We briefly discuss some deployment 
considerations in Section {\ref{deployment}}. Finally, we conclude in Section {\ref{conclusion}}.

%% file: related.tex
\section{Fingerprinting Background}\label{background}

Human fingerprints, due to their unique nature, are a very popular tool used to identify people in forensic and biometric applications~\cite{Ross20032115,cole2009suspect}. Researchers have long sought to find an equivalent of fingerprints in computer systems by finding characteristics that can help identify an individual device. Such fingerprints exploit variation in both the hardware and software of devices to aid in identification.

As early as 1960, the US government used unique transmission characteristics  to track mobile transmitters~\cite{Langley93}. Later, with the introduction of 
cellular network researchers were able to successfully distinguish transmitters by analyzing the spectral characteristics of the transmitted radio 
signal~\cite{Riezenman2000}. Researchers have suggested using radio-frequency fingerprints to enhance wireless authentication~\cite{Li:2006,Nam2011}, as well as localization~\cite{Patwari:2007}.  Others 
have leveraged the minute manufacturing imperfections in network interface cards (NICs) by analyzing the radio-frequency of the emitted 
signals~\cite{Brik:2008,Gerdes06}. Computer clocks have also been used for fingerprinting: Moon et al. showed that network devices tend to have a unique and constant clock skews~\cite{Moon99}; Kohno et al. exploited this to uniquely distinguish network devices through TCP and ICMP timestamps~\cite{Kohno:2005}.

Software can also serve as a distinguishing feature, as different devices have a different installed software base.  
Researchers have long been exploiting the difference in the protocol stack installed on IEEE 
802.11 compliant devices. Desmond et al.~\cite{Desmond:2008} have looked at distinguishing unique devices over Wireless Local Area Networks (WLANs) 
simply by performing timing analysis on the 802.11 probe request packets. Others have investigated subtle differences in the firmware and device 
drivers running on IEEE 802.11 compliant devices~\cite{Franklin:2006}. 802.11 MAC headers have also been used to uniquely track devices~\cite{Guo05}. 
Moreover, there are well-known open source toolkits like Nmap~\cite{nmap} and Xprobe~\cite{xprobe} that can remotely fingerprint an operating 
system by analyzing unique responses from the TCP/IP networking stack.

\paragraph{Browser Fingerprinting} A common application of fingerprinting is to track a user across multiple visits to a website, or a collection of sites. Traditionally, this was done with the aid of cookies explicitly stored by the browser. However, privacy concerns have prompted web browsers to implement features that clear the cookie store, as well as private browsing modes that do not store cookies long-term. This has prompted site operators to develop other means of uniquely identifying and tracking users. Eckersley's Panopticon project showed that many browsers can be uniquely identified by enumerating installed fonts and other browser characteristics, easily accessible via JavaScript~\cite{Eckersley:2010}. A more advanced technique uses HTML5 canvas elements to fingerprint the fonts and rendering engines used by the browser~\cite{canvas-fingerprint}. Others have proposed the use of 
performance benchmarks for differentiating between JavaScript engines~\cite{MBYS11}. Lastly, browsing history can to used to profile and track online 
users~\cite{olejnik:hal-00747841}. Numerous studies have found evidence of these and other techniques being used in the wild~\cite{Acar:2013,Acar:2014,nikiforakis:2012}. 
A number of countermeasures to these techniques exist; typically they disable or restrict the ability of a website to probe the characteristics of a web browser. We expect that smartphones are less susceptible to browser fingerprinting due to a more integrated hardware and software base resulting in less variability, though we are unaware of an exploration of smartphone browser fingerprinting. 

\paragraph{Sensor Fingerprinting} Smartphones do, however, possess an array of sensors that can be used to fingerprint them. Two studies have looked 
at fingerprinting smartphone microphones and speakers~\cite{Das:2014,Zhou:2014}. These techniques, however, require access to the microphone, which is 
typically controlled with a separate permission due to the obvious privacy concerns with the ability to capture audio. Bojinov et 
al.~\cite{BojinovMNB14} additionally consider using accelerometers, which are not considered sensitive and do not require a separate permission. Their 
techniques, however, rely on having the user perform a calibration of the accelerometer (see~\ref{sec:calibration}), the parameters of which are 
used to distinguish phones. Dey et al.~\cite{accelprint} apply machine learning techniques to create an accelerometer fingerprint, but they require 
the vibration motor to be active to stimulate the accelerometer sensor; in the absence of stimulation, they report an average precision and recall of 
only 65\%. 

In contrast, our work studies phones that are in a natural web-browsing setting, either in a user's hand or resting on a flat surface. Additionally, we consider the simultaneous use of both accelerometer and gyroscope to produce a more accurate fingerprint. Inspired by prior work that uses the gyroscope to recover audio signals~\cite{Michalevsky:2014}, we also stimulate the gyroscope with an inaudible tone. Finally, we propose and evaluate several countermeasures to reduce fingerprinting accuracy without entirely blocking access to the motion sensors.

%% file: sensors.tex
\section{A Closer Look at Motion Sensors}{\label{sensors}}
In this section we briefly take a closer look at motion sensors like accelerometer and gyroscope that are embedded in today's smartphones. This will 
provide an understanding of how they can be used to uniquely fingerprint smartphones. Accelerometer and gyroscope sensors in modern smartphones are 
based on Micro Electro Mechanical Systems (MEMS). STMicroelectronics~\cite{STMicroelectronics} and InvenSense~\cite{invensense} are among the top 
vendors supplying MEMS-based accelerometer and gyroscope sensor to different smartphone manufacturers~\cite{MEMSmarket}. Traditionally, 
Apple~\cite{iphone4,iphone5}\footnote{However, iphone 6 has been reported to use motion sensors produced by InvenSense.} and 
Samsung~\cite{galaxys3,galaxys4} favor using STMicroelectronics motion sensors, while Google~\cite{nexus4,nexus5} tends to use InvenSense sensors.

\subsection{Accelerometer}
\begin{figure}[!h]
\centering
\includegraphics[width=0.85\columnwidth]{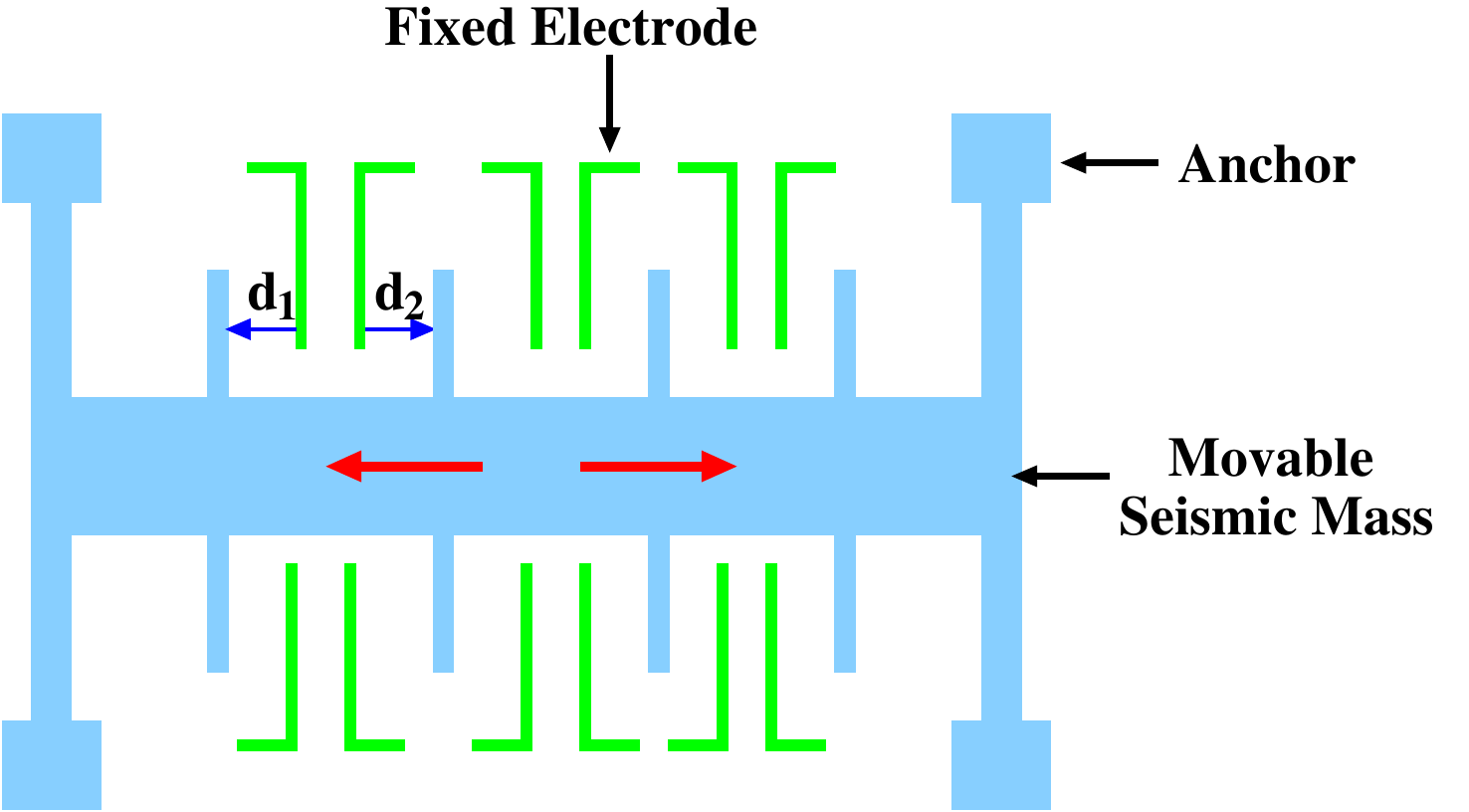}
\caption{Internal architecture of a MEMS accelerometer. Differential capacitance is proportional to the applied acceleration.} 
\label{accel}
\end{figure}

{\label{Accelerometer}}
Accelerometer is a device that measures proper acceleration. Proper acceleration is different from coordinate acceleration (linear acceleration) as 
it measures the \emph{g-force}. For example, an accelerometer at rest on a surface will measure an acceleration of $g=9.81 ms^{-2}$ straight upwards, 
while for a free falling object it will measure an acceleration of zero. MEMS-based accelerometers are based on differential 
capacitors~\cite{memsaccelerometer}. Figure~\ref{accel} shows the internal architecture of a MEMS-based accelerometer. As we can we there are several 
pairs of fixed electrodes and a movable seismic mass. Under zero force the distances $d_1$ and $d_2$ are equal and as a result the two capacitors are 
equal, but a change in force will cause the movable seismic mass to shift closer to one of the fixed electrodes (i.e., $d_1 \neq d_2$) causing a 
change in the generated capacitance. This difference in capacitance is detected and amplified to produce a voltage proportional to the acceleration. 
The slightest gap between the structural electrodes, introduced during the manufacturing process, can cause a change in the capacitance. Also the 
flexibility of the seismic mass can be slightly different from one chip to another. This form of minute imprecisions in the electro-mechanical 
structure induce subtle imperfections in accelerometer chips. 

\subsection{Gyroscope}
\begin{figure}[!htb]
\centering
\begin{tabular}{cc}
\epsfig{file=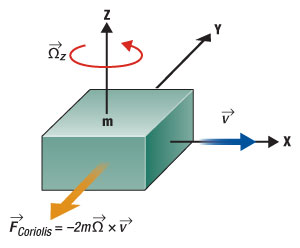,width=0.35\columnwidth,clip=}&\epsfig{file=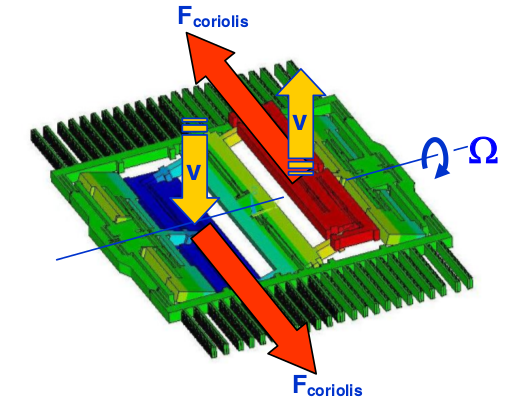,width=0.6\columnwidth,clip=}
\end{tabular}
\caption{MEMS-based gyros use {\em Coriolis force} to compute angular velocity. The Coriolis force induces change in capacitance which is proportional 
to the angular velocity.} 
\label{gyro}
\end{figure}

{\label{Gyroscope}}
Gyroscope measures the rate of rotation (in $rads^{-1}$) along the device's three axes. MEMS-based gyroscopes use the Coriolis effect to measure 
the angular rate. Whenever an angular velocity of $\omega$ is exerted on a moving mass of weight $m$ and velocity $\hat{v}$, the object experiences a 
Coriolis force in a direction perpendicular to the rotation axis and to the velocity of the moving object (as shown in figure~\ref{gyro}). The 
Coriolis force is calculated by the following equation $F=2m\hat{v}\times\omega$. Generally, the angular rate ($\omega$) is measured by sensing the 
magnitude of the Coriolis force exerted on a vibrating proof-mass within the gyro~\cite{memsgyro2,memsgyro3,memsgyro1}. The Coriolis force is sensed 
by a capacitive sensing structure where a change in the vibration of the proof-mass causes a change in capacitance  which is then converted into a 
voltage signal by the internal circuitry. Again the slightest imperfection in the electro-mechanical structure will introduce idiosyncrasies across 
chips. 

%% file: features.tex
\section{Features and Classification Algorithms}{\label{feature_algo}}
In this section we briefly describe the data pre-processing procedure and the features used in generating a device fingerprint. We also discuss the 
classification algorithms and metrics used in our evaluation.

\begin{table*}[tbp]
\centering \caption{Explored acoustic features} \resizebox{16cm}{!}{
\begin{tabular}{|c|c|c|c|}
\hline
\#&Domain&Feature&Description\\
\hline
1&\multirow{10}{*}{Time}&Mean&The arithmetic mean of the signal strength at different timestamps\\
\cline{3-4}
2&&Standard Deviation&Standard deviation of the signal strength\\
\cline{3-4}
3&&Average Deviation&Average deviation from mean\\
\cline{3-4}
4&&Skewness&Measure of asymmetry about mean\\
\cline{3-4}
5&&Kurtosis&Measure of the flatness or spikiness of a distribution\\
\cline{3-4}
6&&RMS&Square root of the arithmetic mean of the squares of the signal strength at various timestamps\\
\cline{3-4}
7&&Max&Maximum signal strength\\
\cline{3-4}
8&&Min&Minimum signal strength\\
\cline{3-4}
9&&ZCR&The rate at which the signal changes sign from positive to negative or back\\
\cline{3-4}
10&&Non-Negative count&Number of non-negative values\\
\hline
11&\multirow{15}{*}{Frequency}&Spectral Centroid&Represents the center of mass of a spectral power distribution\\
\cline{3-4}
12&&Spectral Spread&Defines the dispersion of the spectrum around its centroid\\
\cline{3-4}
13&&Spectral Skewness&Represents the coefficient of skewness of a spectrum\\
\cline{3-4}
14&&Spectral Kurtosis&Measure of the flatness or spikiness of a distribution relative to a normal distribution\\
\cline{3-4}
15&&Spectral Entropy&Captures the peaks of a spectrum and their locations\\
\cline{3-4}
16&&Spectral Flatness&Measures how energy is spread across the spectrum\\
\cline{3-4}
17&&Spectral Brightness&Amount of spectral energy corresponding to frequencies higher than a given cut-off threshold\\
\cline{3-4}
18&&Spectral Rolloff&Defines the frequency below which 85\% of the distribution magnitude is concentrated\\
\cline{3-4}
19&&Spectral Roughness&Average of all the dissonance between all possible pairs of peaks in a spectrum\\
\cline{3-4}
20&&Spectral Irregularity&Measures the degree of variation of the successive peaks of a spectrum\\
\cline{3-4}
21&&Spectral RMS&Square root of the arithmetic mean of the squares of the signal strength at various frequencies\\
\cline{3-4}
22&&Low-Energy-Rate&The percentage of frames with RMS power less than the average RMS power for the whole signal\\
\cline{3-4}
23&&Spectral flux&Measure of how quickly the power spectrum of a signal changes\\
\cline{3-4}
24&&Spectral Attack Time&Average rise time to spectral peaks\\
\cline{3-4}
25&&Spectral Attack Slope&Average slope to spectral peaks\\
\hline
\end{tabular}}
\label{allfeatures}
\end{table*}

\subsection{Data Preprocessing}{\label{preprocess}}
Data from motion sensors can be thought of as a stream of timestamped real values. For both accelerometer and gyroscope we obtain values along 
 three axes. So, for a given timestamp, $t$, we have two vectors of the following form: $\vec{a}(t)=(a_x,a_y,a_z)$ and 
$\vec{\omega}(t)=(\omega_x,\omega_y,\omega_z)$. The accelerometer values include gravity, i.e., when the device is stationary lying flat on top of 
a surface we get a value of $9.81 ms^{-2}$ along the $z$-axis. 
We convert the acceleration vector into a scalar by taking its magnitude: $|\vec{a}(t)|=\sqrt{a_x^2+a_y^2+a_z^2}$. This technique discards some information, but has the advantage of making the accelerometer data independent of device orientation; e.g., if the device is stationary the acceleration magnitude will always be around $9.81 ms^{-2}$, whereas the reading on each individual axis will vary greatly (by +/- $1g$) depending on how the device is held. For the gyroscope we consider data from each axis as a separate stream, since there is no corresponding baseline rotational acceleration. In other words, if the device is stationary the rotation rate across 
all three axes should be close to $0$ $\mathtt{rad}s^{-1}$, irrespective of the orientation of the device. Thus, our model considers four 
streams of sensor data in the form of $\{|\vec{a}(t)|,\omega_x(t),\omega_y(t),\omega_z(t)\}$.

For all data streams, we also look at frequency domain characteristics. But since the browser, running as one of many applications inside the 
phone,  makes API calls to collect sensor data the OS might not necessarily respond in a synchronized manner\footnote{Depending on the load and 
other applications running, OS might prioritize such API calls differently.}. This results in non-equally spaced data points. We, therefore, use 
cubic-spline interpolation~\cite{cubicspline} to construct new data points such that $\{|\vec{a}(t)|,\omega_x(t),\omega_y(t),\omega_z(t)\}$ become
equally-spaced.

\subsection{Temporal and Spectral Features}{\label{features}}
To summarize the characteristics of a sensor data stream, we explore a total of 25 features consisting of 10 temporal and 15 spectral 
features (listed in Table~\ref{allfeatures}). All of these features have been well documented by researchers in the past. A detailed description of 
each feature is available in Appendix~\ref{appendix_features}.

\subsection{Classification Algorithms and Metrics}{\label{classification-algo}}
\paragraphb{Classification Algorithms:}
Once we have features extracted from the sensor data, we use supervised learning to identify the source sensor. Any supervised learning 
classifier has two main phases: training phase and testing phase. During training, features from all smartphones (i.e., labeled data) are used to 
train the classifier. In the test phase, the classifier predicts the the most probable class for a given (unseen) feature vector. We evaluate the 
performance of the following supervised classifiers --- Support Vector Machine (SVM), Naive-Bayes classifier, Multiclass Decision Tree, 
k-Nearest Neighbor (k-NN), Quadratic Discriminant Analysis (QDA) classifier and Bagged Decision Trees (Matlab's Treebagger 
model)~\cite{matlabalgos}. We found that in general ensemble based approaches like Bagged Decision Trees outperform the other classifiers. We 
report the maximum achievable accuracies from these classifiers in the evaluation Section~\ref{evaluation}.

\paragraphb{Evaluation metrics:}
For evaluation metric we use standard multi-class classification metrics like---\emph{precision}, \emph{recall}, and 
\emph{F-score}~\cite{Sokolova2009427}---in our evaluation. Assuming there are $n$ classes, we first compute the true positive 
($TP$) rate for each class, i.e., the number of traces from the class that are classified correctly. Similarly, we compute the false positive ($FP$) 
and false negative ($FN$) as the number of wrongly accepted and wrongly rejected traces, respectively, for each class $i$ ($1\leq i\leq n$). We then 
compute precision, recall, and the F-score for each class using the following equations:\nolinebreak
\begin{align}
\mbox{Precision,  } Pr_i &= {TP_i}/(TP_i+FP_i)\\
\mbox{Recall,  } Re_i &= {TP_i}/(TP_i+FN_i)\\
\mbox{F-Score,  } \mathit{F}_i &= ({2\times Pr_i\times Re_i})/(Pr_i+Re_i)
\end{align}
The F-score is the harmonic mean of precision and recall; it provides a good measure of overall classification performance, since 
precision and recall represent a trade-off: a more conservative classifier that rejects more instances will have higher precision but lower recall, 
and vice-versa. To obtain the overall performance of the system we compute average values in the following way:\nolinebreak
\begin{align}
\mbox{Avg. Precision,  } \mathit{AvgPr} &= \frac{\sum_{i=1}^{n}Pr_i}{n}\\
\mbox{Avg. Recall,  } \mathit{AvgRe} &= \frac{\sum_{i=1}^{n}Re_i}{n}\\
\mbox{Avg. F-Score,  } \mathit{AvgF} &= \frac{2\times AvgPr\times AvgRe}{AvgPr+AvgRe}
\end{align}

%% file: evaluation.tex
\section{Fingerprinting Evaluation}{\label{evaluation}}
In this section we first describe our experimental setup (Section~\ref{expsetup}). We then explore features with the aim to determine the minimal 
subset of features required to maximize classification accuracy (Section~\ref{feature_exploration}). Lastly, we evaluate our fingerprinting approach 
under a controlled lab setting (Section~\ref{labsetting}), an uncontrolled real-world setting (Section~\ref{publicsetting}) and a combination of both 
settings (Section~\ref{combosetting}).

\subsection{Experimental Setup}{\label{expsetup}}
Our experimental setup consists of developing our own web page to collect sensor 
data\footnote{\url{http://datarepo.cs.illinois.edu/SensorFingerprinting.html}}. We use a simple Javascript to access accelerometer and 
gyroscope data (sample code snippet is provided in Appendix~\ref{appendix_code}). However, since we collect data through the browser the maximum 
obtainable sampling frequency is lower than available hardware sampling frequency (restricted by the underlying OS). Table~\ref{samplefreq} 
summarizes the sampling frequencies obtained from the top 5 mobile 
browsers~\cite{topmobilebrowser}\footnote{We computed the average time it took to obtain 100 
samples. Sample website available at \url{http://datarepo.cs.illinois.edu/SamplingFreq.html}}. We use a Samsung Galaxy S3 and iPhone 5 to test 
the sampling frequency of the different browsers. Table~\ref{samplefreq} also highlights the motion sensors that are accessible from the different 
browsers. We see that Chrome provides the best sampling frequency while the default Android browser is the most restrictive browser in terms of not 
only sampling frequency but also access to different motion sensors. However, Chrome being the most popular mobile browser~\cite{browsertrend}, we 
collect data using the Chrome browser.

\begin{table}[!h]
\centering \caption{Sampling frequency from different browsers} \resizebox{8cm}{!}{
\begin{tabular}{|c|c|c|c|}
\hline
\multirow{2}{*}{OS}&\multirow{2}{*}{Browser}&Sampling&Accessible\\
&&Frequency ($\sim$Hz)&Sensors$^\ast$\\
\hline
\multirow{5}{*}{Android 4.4}&Chrome&100&A,G\\
\cline{2-4}
&Android&20&A\\
\cline{2-4}
&Opera&40&A,G\\
\cline{2-4}
&UC Browser&20&A,G\\
\cline{2-4}
&Standalone App~\cite{androidSDK}&200&A,G\\
\hline
\multirow{2}{*}{iOS 8.1.3}&Safari&40&A,G\\
\cline{2-4}
&Standalone App~\cite{appleSDK}&100&A,G\\
\hline
\end{tabular}}
{\footnotesize $\ast$ here \textbf{'A'} means accelerometer and \textbf{'G'} refers to gyroscope}
\label{samplefreq}
\end{table}

We start off our data collection from 30 lab-smartphones. Table~\ref{phones} lists the distribution of the different smartphones from 
which we collect sensor data. Now, as gyroscopes react to audio stimulation we collect data under three different background audio settings: no audio, an inaudible 20\,kHz sine wave, or a popular song playing. In the latter two scenarios, the corresponding audio file plays in the background of the 
browser while data is being collected. Under each setting we collect 10 samples where each sample is about 5 to 8 seconds worth of data. Now, since 
our fingerprinting approach aims to capture the inherent imperfections of motion sensors, we need to keep the sensors stationary while collecting 
data. Therefore, by default, we have the phone placed flat on a surface while data is being collected, unless explicitly stated otherwise. We, 
however, do test our approach for the scenario where the user is holding the smartphone in his/her hand while sitting down.

For training and testing the classifiers we \emph{randomly} split the dataset in such a way that 50\% of data from each device goes to the training 
set while the remaining 50\% goes to the test set. To prevent any bias in the selection of the training and testing set, we randomize the training and 
testing set 10 times and report the average F-score. We also compute the 95\% confidence interval, but we found it to be less than 1\% and 
therefore, do not report it in the rest of the paper. For analyzing and matching fingerprints we use a desktop machine with an Intel i7-2600 3.4GHz 
processor with 12GiB RAM. We found that the average time required to match a new fingerprint was around 10--100\,ms.

\begin{table}[!h]
\centering \caption{Types of phones used} \resizebox{5cm}{!}{
\begin{tabular}{|c|c|c|}
\hline
Maker&Model&Quantity\\
\hline
\multirow{2}{*}{Apple}&iPhone 5&4\\
\cline{2-3}
&iPhone 5s&3\\
\hline
\multirow{3}{*}{Samsung}&Nexus S&14\\
\hline
&Galaxy S3&4\\
\cline{2-3}
&Galaxy S4&5\\
\hline
\multicolumn{2}{|c|}{Total}&30\\
\hline
\end{tabular}}
\label{phones}
\end{table}


\begin{figure*}[!htb]
\centering
\begin{tabular}{ccc}
\epsfig{file=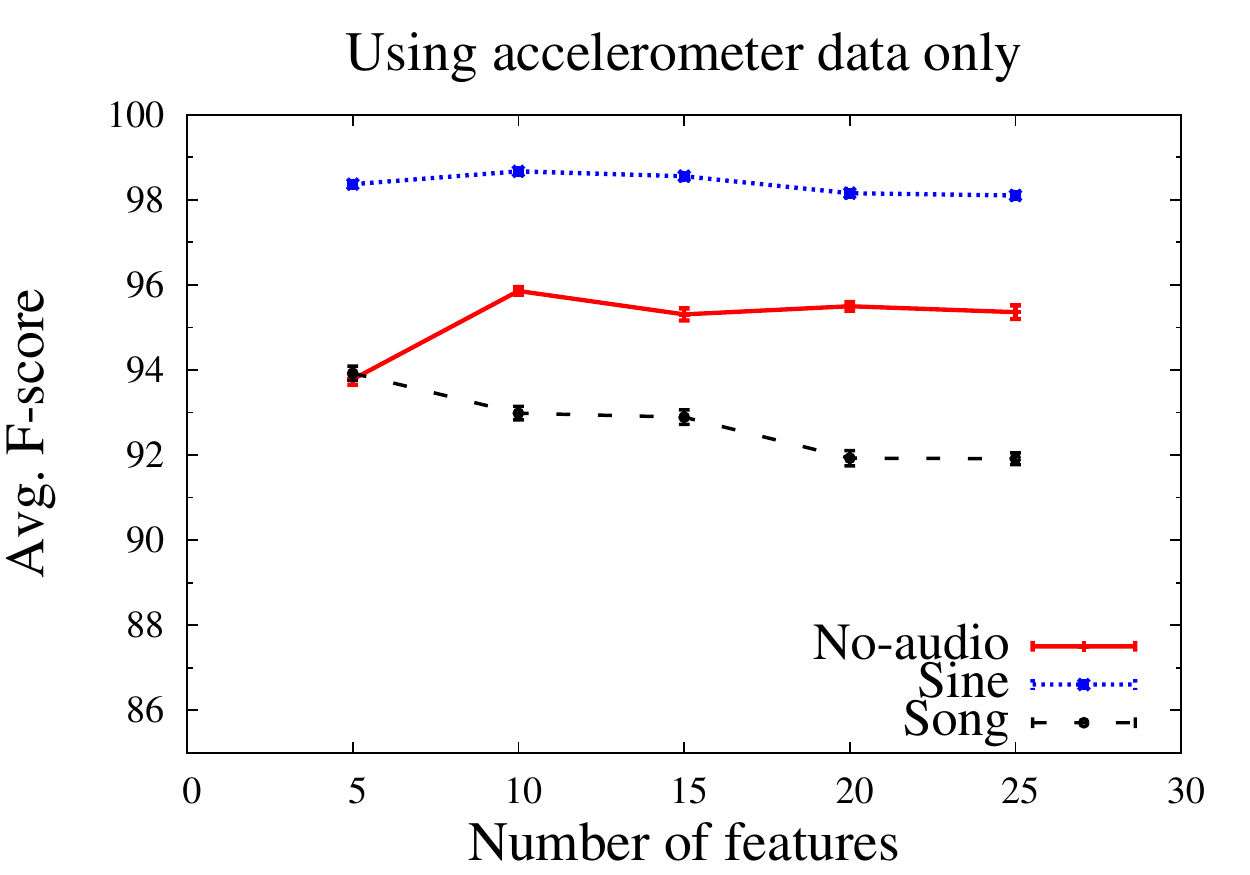,width=0.32\linewidth,clip=}&\epsfig{file=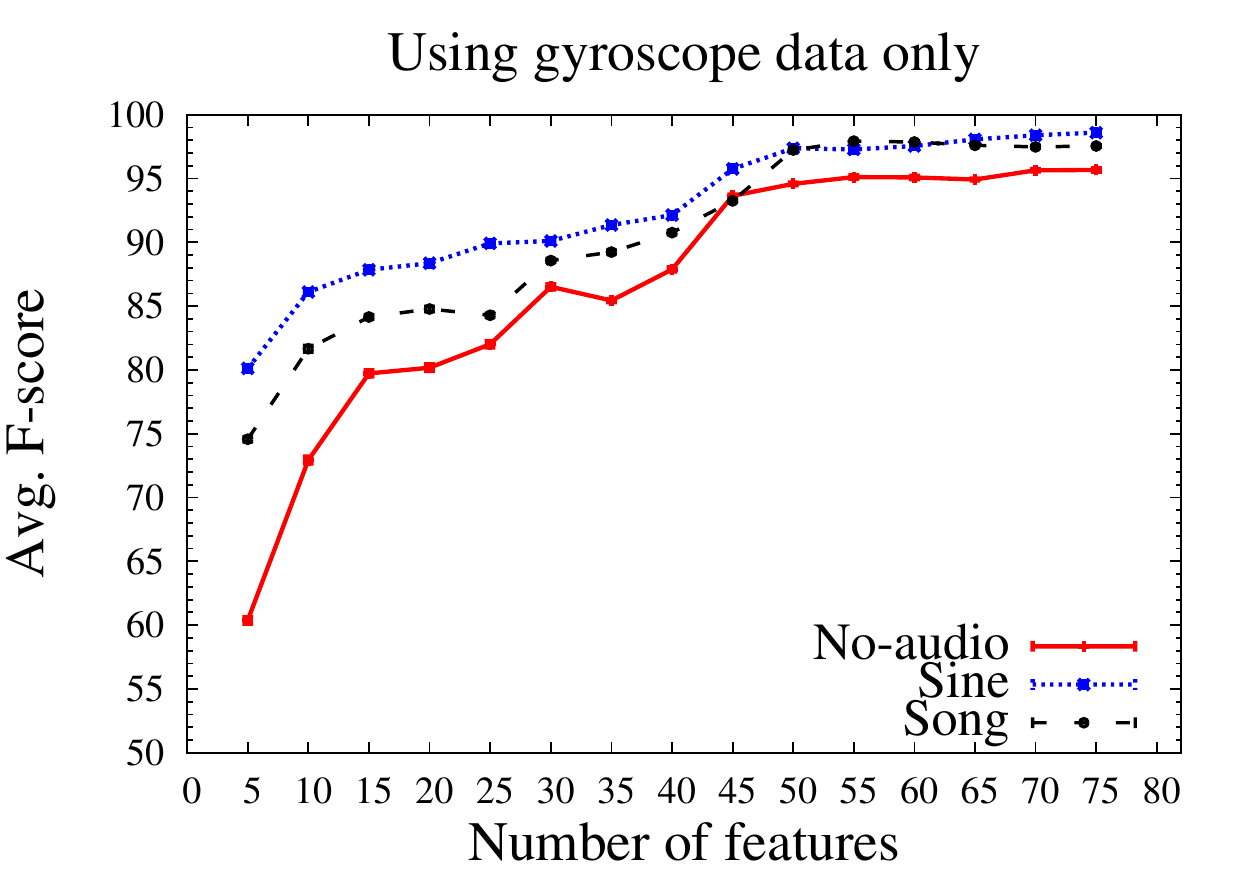,width=0.32\linewidth,clip=}&\epsfig{
file=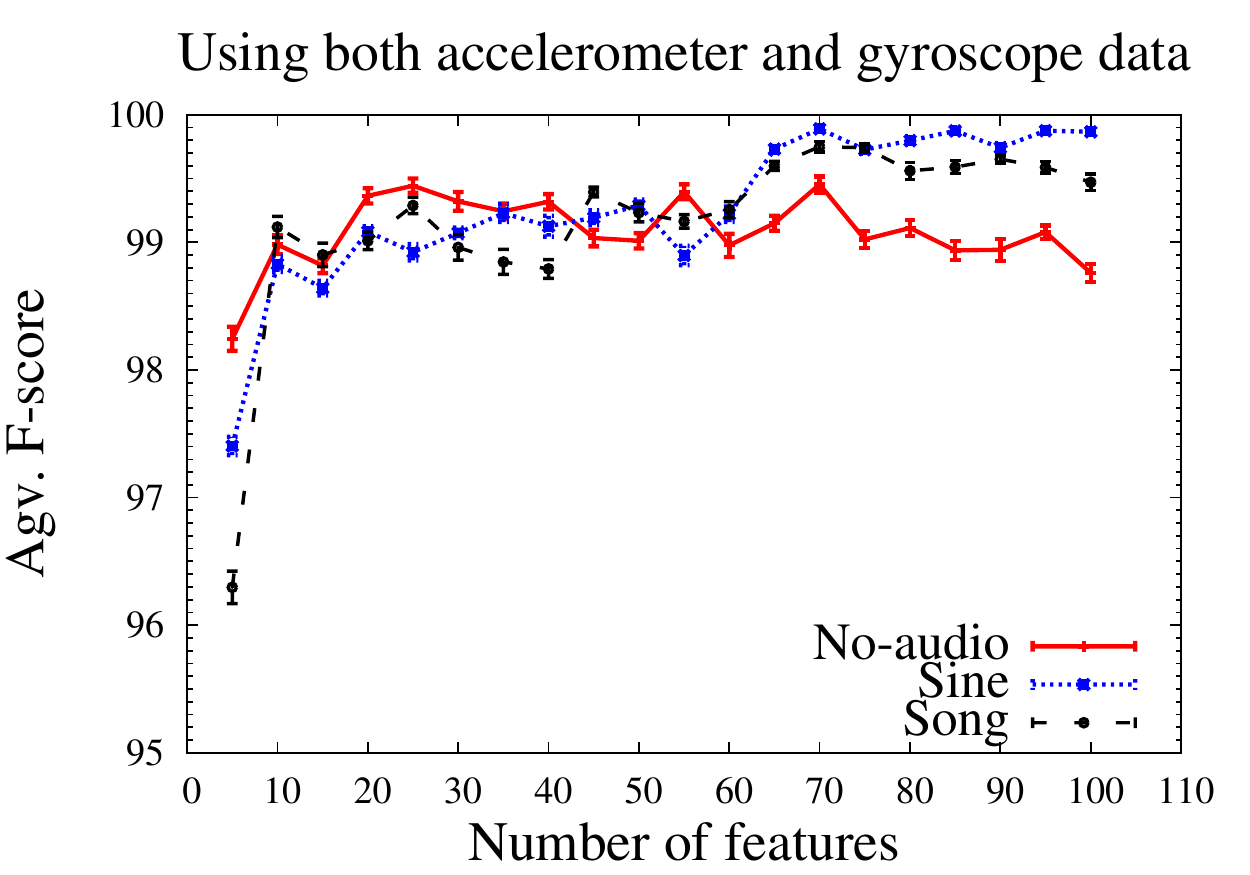,width=0.32\linewidth,clip=}
\end{tabular}
\caption{Exploring the number optimal features for different sensors. For a) accelerometer more than top 10 features leads to diminished returns, b) 
gyroscope all 75 features contribute to obtaining improved accuracy, c) combined sensor data more than 70 features leads to diminished returns.} 
\label{featureexploration}
\end{figure*}

\subsection{Feature Exploration and Selection}{\label{feature_exploration}}
At first glance, it might seem that using all features at our disposal to identify the device is the optimal strategy. However, including too 
many features can worsen performance in practice, due to their varying accuracies and potentially-conflicting signatures. We, therefore, explore all 
the features and determine the subset of features that optimize our fingerprinting accuracy. For temporal features, no transformation of the data 
stream is required, but for spectral features we first convert the non-equally spaced data stream into a fixed-spaced data stream using cubic 
spline interpolation. We interpolate at a sampling rate of 8kHz\footnote{Although up-sampling the signal from $\sim$100 Hz to 8 kHz does not
increase the accuracy of the signal, it does make direct application of standard signal processing tools more convenient.}. Then, we use the 
following signal analytic tools and modules: \emph{MIRtoolbox}~\cite{mirtoolbox} and \emph{Libxtract}~\cite{libxtract} to extract spectral features. 
We, next look at feature selection where we explore different combinations of features to maximize our fingerprinting accuracy. We use the FEAST 
toolbox~\cite{pocbro14} and utilize the \emph{Joint Mutual Information} criterion (JMI criterion is known to provide the best tradeoff in terms of 
accuracy, stability, and flexibility with small data samples~\cite{Brown:2012}) for ranking the features.

Figure~\ref{featureexploration} shows the results of our feature exploration for the 30 lab-smartphones. We see that when using only accelerometer 
data the F-score seems to flatten after considering the top 10 features. For gyroscope data we see that using all the 75 features (25 per data 
stream) obtains the best result. And finally when we combine both accelerometer and gyroscope features, the top 70 features (from a total of 100 
features) seems to provide the best fingerprinting accuracy. Among these top 70 features we found that 21 of them came from accelerometer features and 
the remaining 59 came from gyroscope features. In terms of the distribution between temporal and spectral features, we found that spectral features 
dominated with 44 of the top 70 features being spectral features. We use these subset of features in all our latter evaluations.

\subsection{Results From Lab Setting}\label{labsetting}
\begin{figure}[!htb]
\centering
\begin{tabular}{c}
\epsfig{file=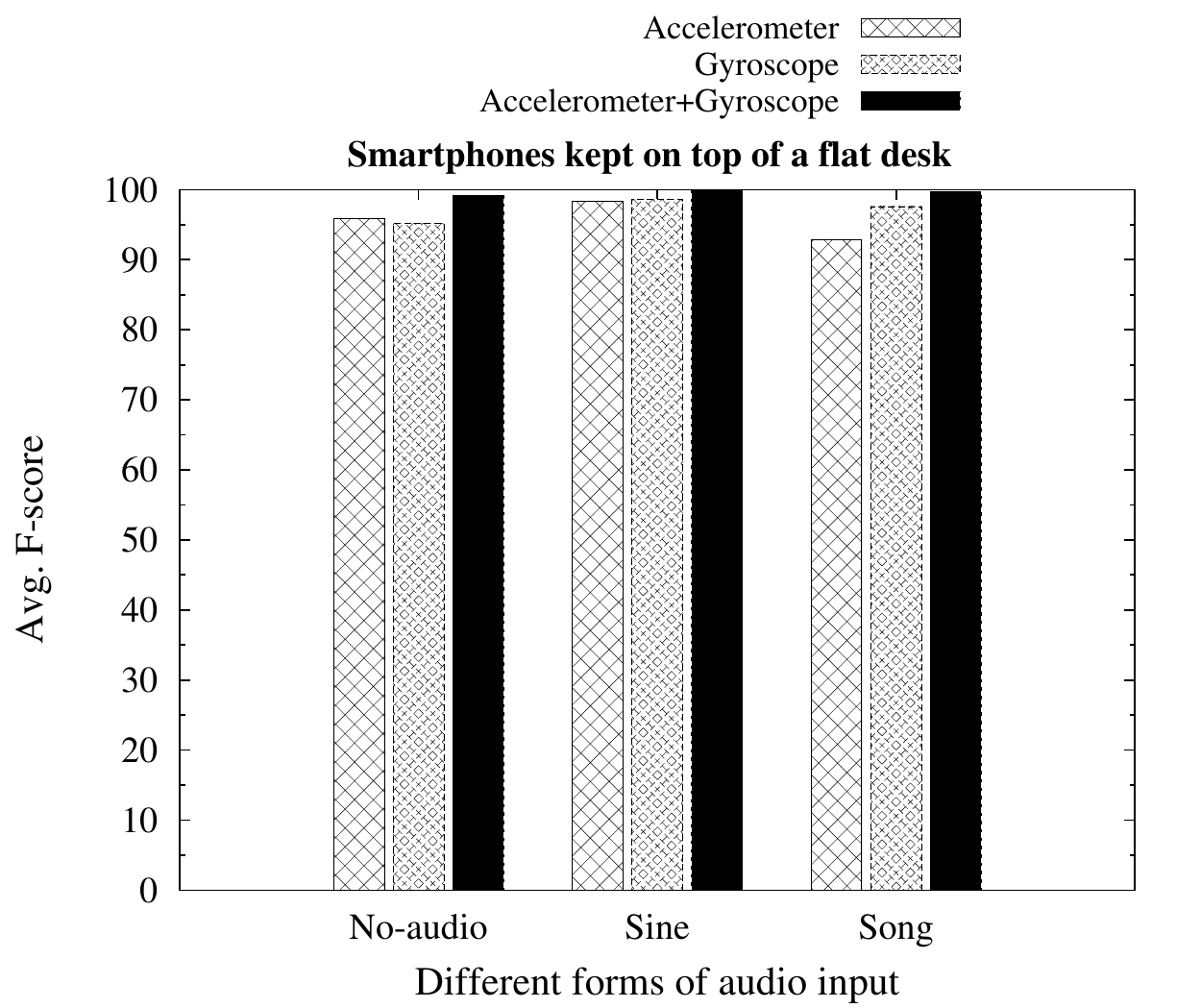,width=0.9\columnwidth,clip=}\\
(a)\\
\epsfig{file=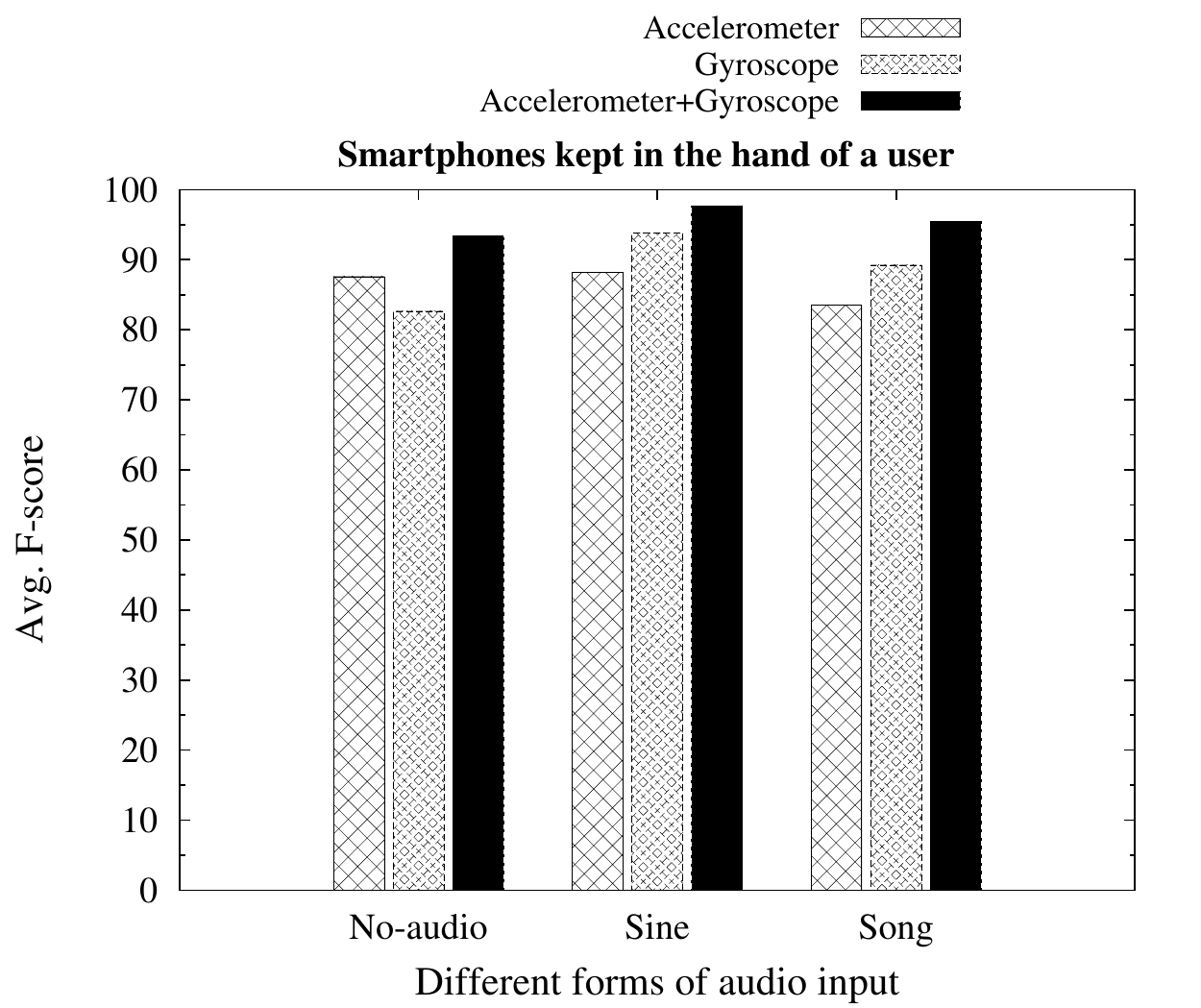,width=0.9\columnwidth,clip=}\\
(b)
\end{tabular}
\caption{Average F-score for different forms of audio stimulation under lab setting. For a) smartphones are kept \emph{on top of a desk} while 
collecting sensor data, b) smartphones are kept \emph{in the hand of the user} while collecting sensor data.} 
\label{lab_ondesk_inhand}
\end{figure}

First, we look at fingerprinting smartphones under lab environment to demonstrate the basic viability of the attack. For this purpose
we keep smartphones stationary on top of a flat surface. Figure~\ref{lab_ondesk_inhand}(a) summarizes 
our results. We see that we can almost correctly identify all 30 smartphones under all three scenarios by combining the accelerometer and gyroscope 
features. While the benefit of the background audio stimulation is not visible from the figure, we will later on show that audio stimulation do in 
fact enhance fingerprinting accuracy under countermeasure techniques like calibration and obfuscation (more in Section~\ref{defense}). Overall these 
results indicate that it is indeed possible to fingerprint smartphones through motion sensors.

\subsection{Results From Public Setting}\label{publicsetting}
\begin{figure}
\centering
\includegraphics[width=1.0\columnwidth]{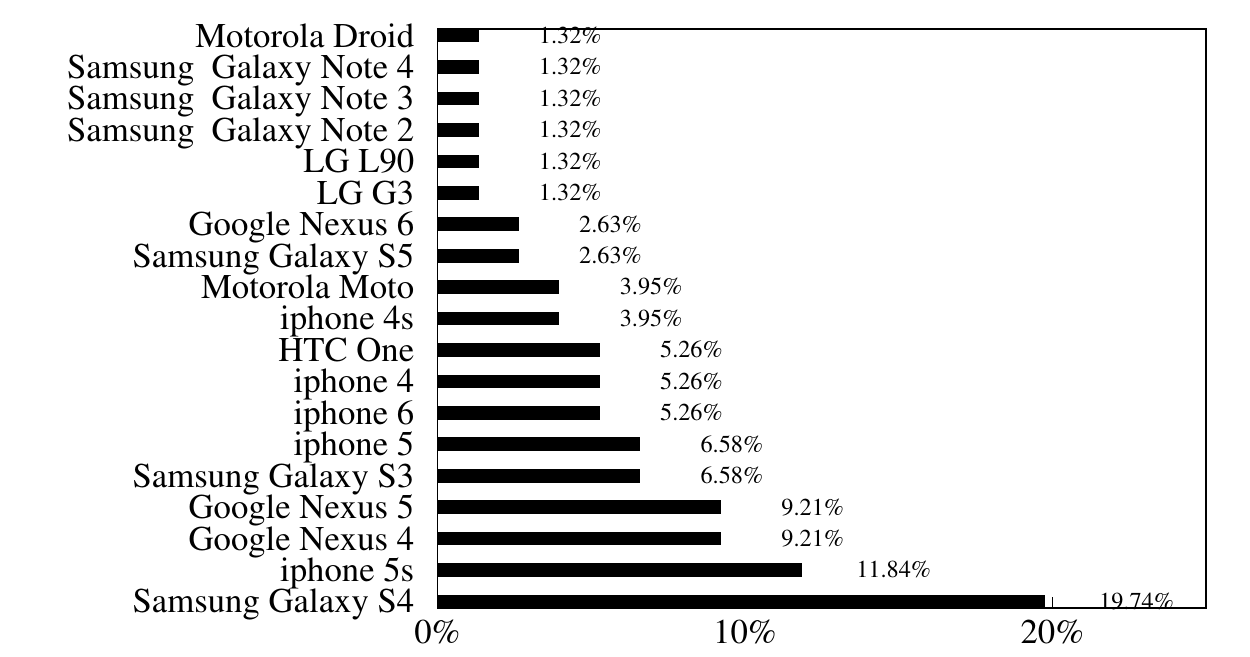}
\caption{Distribution of participant device model.}
\label{devices}
\end{figure}

After gaining promising results from our relatively small-scale lab setting, we set out to expand our data collection process to real-world 
public setting. We invited people to voluntarily participate in our study by visiting our web 
page\footnote{Available at \url{http://datarepo.cs.illinois.edu/DataCollectionHowPlaced.html}. Screenshots of our data collection webpage is provided 
in Appendix~\ref{appendix_screenshot}. We obtained approval from our Institutional Research Board (IRB) to perform the data collection.} and following 
a few simple steps to provide us with sensor data. We recruited participants through email 
and online social networks. We asked participants to provide data under two settings: no-audio setting and the inaudible sine-wave setting. (We avoid the background song to make the experience less bothersome for the user and more realistic.) Each setting collected sensor data for about one minute, requiring a total of two minutes of participation. (We did 
not ask participants to provide data under all three settings because it would require more time which could potentially discourage participants to 
not fully finish their task.) On average, we had around 10 samples per setting per device. Our data-gathering web page plants a cookie in the form of 
a large random number (acting as a unique ID) in the user's browser, which makes it possible to correlate data points coming from the same device. Over 
the course of two weeks we received data from a total of 76 devices. However, some participants did not follow all the steps and as a result we were 
able to use only 63 of the 76 submissions. Figure~\ref{devices} shows the distribution of the different devices that participated in our study.

\begin{figure}
\centering
\includegraphics[width=0.9\columnwidth]{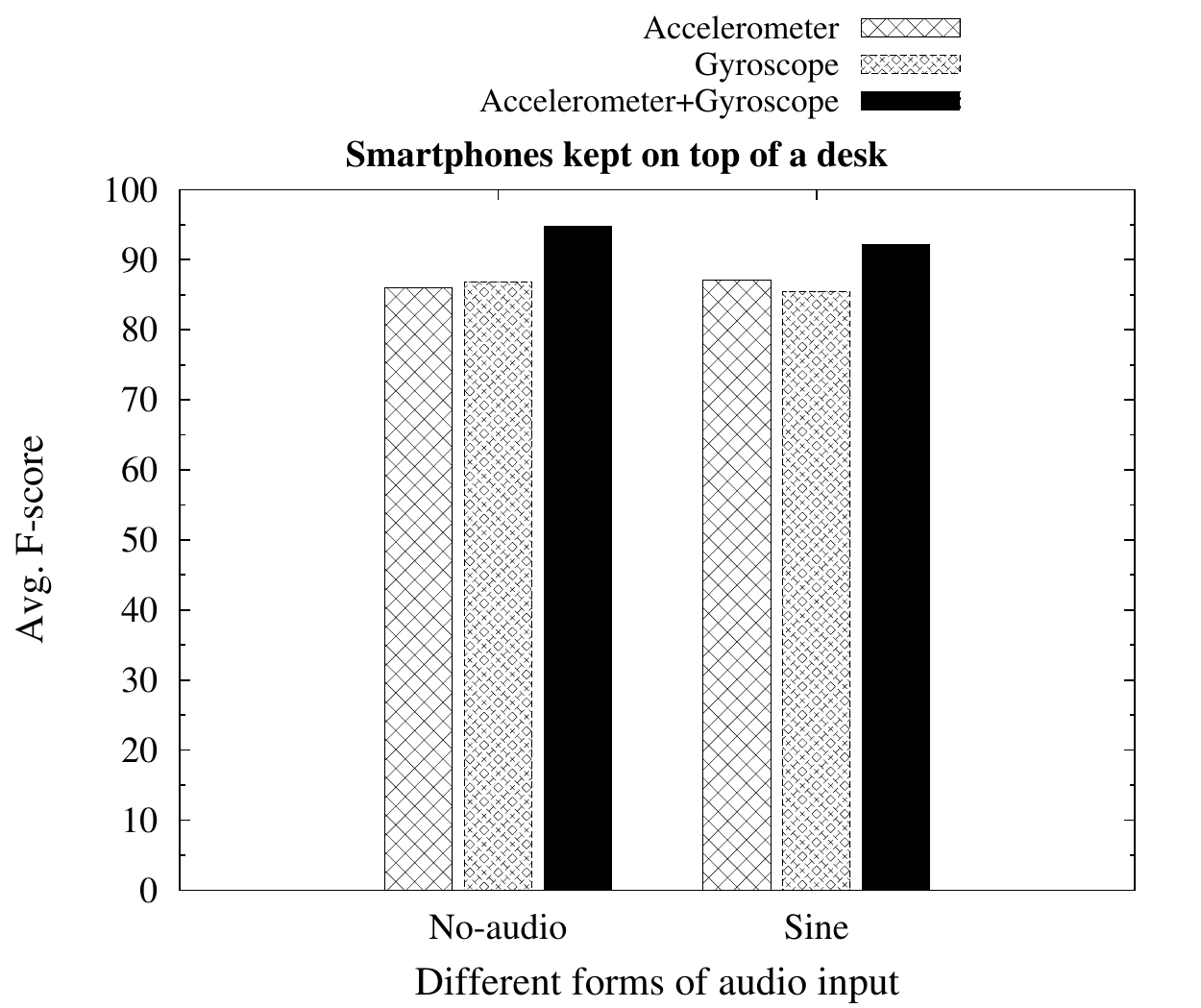}
\caption{Average F-score for different forms of audio stimulation under public setting. Results obtained for 63 public smartphones 
where users were told to keep the smartphone \emph{on top of a desk} while collecting sensor data.}
\label{public_plain}
\end{figure}

Next, we apply our fingerprinting approach on the public data set. Figure~\ref{public_plain} shows our findings. Compared to the results from our lab 
setting, we see a slight decrease in F-score but even then we were able to obtain an F-score of $\sim94\%$. Again, the benefit of the audio 
stimulation is not evident from these results, however, their benefits will become more visible in the later sections when we discuss about 
countermeasure techniques. 

\subsection{Results From Combined Setting}\label{combosetting}
\begin{figure}[!htb]
\centering
\includegraphics[width=0.9\columnwidth]{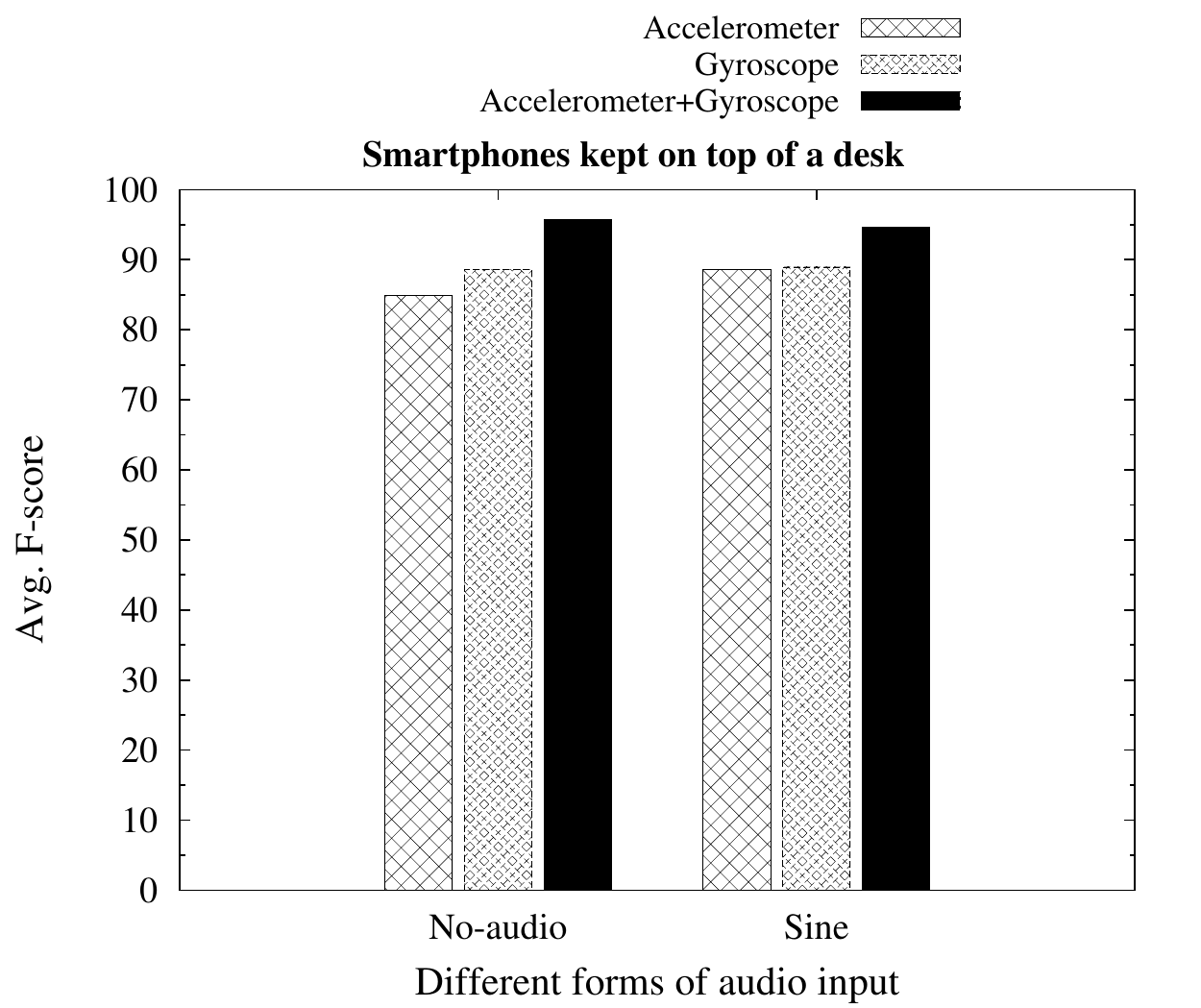}
\caption{Average F-score for different forms of audio stimulation. Results are obtained by combining the publicly collected data with our lab data 
giving us a total of 93 devices. All the smartphones were kept \emph{on top of a desk} while collecting sensor data.}
\label{combo_plain}
\end{figure}

Finally, we combine our lab data with the publicly collected data to give us a combined dataset containing 93 different smartphones. We
apply the same set of evaluations on this combined dataset. Figure~\ref{combo_plain} highlights our findings. Again, we see that combining features 
from both sensors provides the best result. In this case we obtained an F-score of $~\sim96\%$. All these results suggest that smartphones 
can be successfully fingerprinted through motion sensors.

%% file: sensitivity.tex
\subsection{Sensitivity Analysis}{\label{sensitivity}}

\subsubsection{Varying the Number of Devices}

\begin{figure}
\centering
\includegraphics[width=0.9\columnwidth]{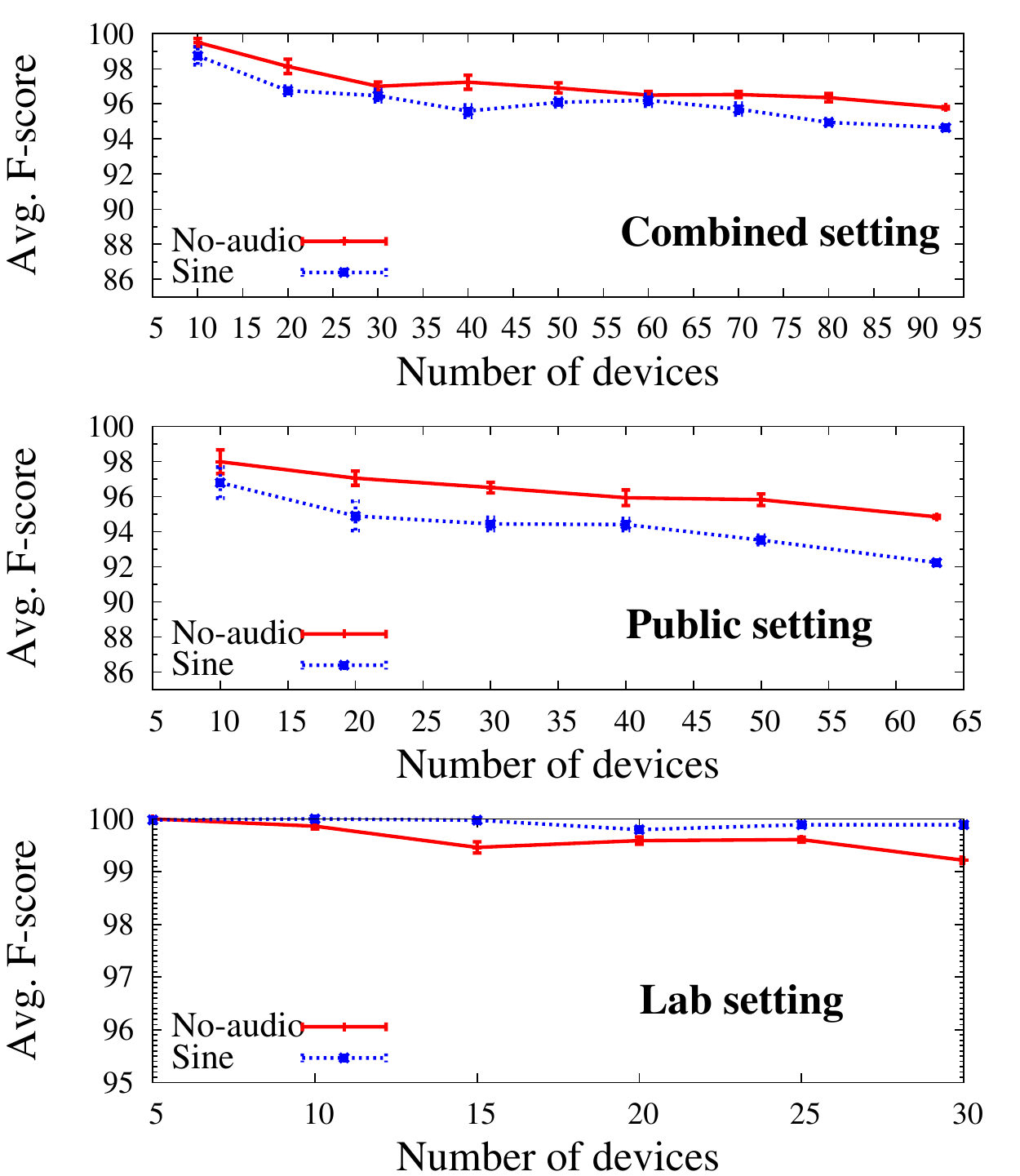}
\caption{Average F-score for different numbers of smartphones. F-score generally tends to decrease slightly as more devices are considered.}
\label{varyphn}
\end{figure}

We evaluate the accuracy of our classifier while varying the number of devices. We pick a subset of $n$ devices in our data set and perform the training and testing steps for this subset. For each value of $n$, we repeat the experiment 10 times, using a different random subset of $n$ devices each time. In this experiment we consider the use of both accelerometer and gyroscope features, since those produce the best performance, and focus on the no audio and sine wave background scenarios. Figure~\ref{varyphn} shows that the F-score generally decreases with large number of devices, which is expected as an increased number of labels makes classification more difficult. Extrapolating from the graph, we expect classification to remain accurate even for significantly larger data sets.

\subsubsection{Varying Training Set Size}

\begin{figure}
\centering
\includegraphics[width=0.9\columnwidth]{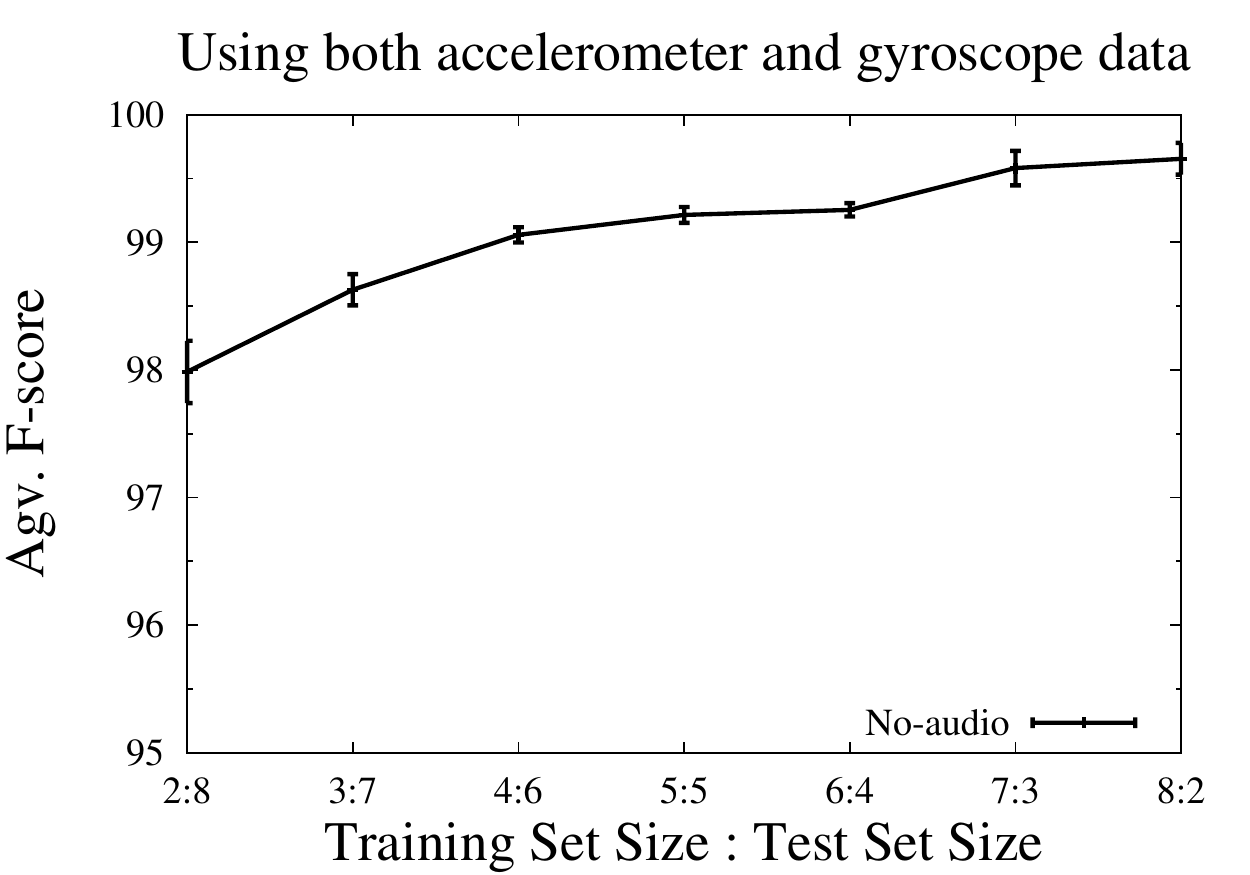}
\caption{Average F-score for different ratio of training and test data. With only two training data we achieved a F-score of $\sim98\%$ }
\label{varyptraining}
\end{figure}

We also consider how varying the training set size impacts the fingerprinting accuracy. For this experiment we vary the ratio of training and testing 
set size. For this experiment we only look at data from our lab setting as some of the devices from our public setting did not have exactly 10 
samples. We also consider the setting where there is no background audio stimulation and use the combined features of accelerometer and gyroscope. 
Figure~\ref{varyptraining} shows our findings.  While an increased training size improves classification accuracy, even with mere two training samples (of a few seconds each) are sufficient to achieve an F-score of $\sim 98$, with increased training set sizes producing an F-score of over 99\%.

%% file: countermeasure.tex
\section{Countermeasures}{\label{defense}}
So far we have focused on showing how easy it is to fingerprint smartphones through motion sensors. We now shift our focus on providing a systematic 
approach to defending against such fingerprinting techniques. We propose two approaches: sensor calibration and data obfuscation.

\subsection{Calibration}
\label{sec:calibration}

Bojinov et al.~\cite{BojinovMNB14} observe that their phones have calibration errors, and use these calibration differences as a mechanism to distinguish between them. In particular, they consider an affine error model: $a^M= g\cdot a + o$, where $a$ is the true acceleration along an axis and $a^M$ is the measured value of the sensor. The two error parameters are the offset $o$ (bias away from 0) and the gain $g$ which magnifies or diminishes the acceleration value.
Our classification uses many features, but we find that the \emph{mean} signal value is the most discriminating feature for each of the sensor streams, which is closely related to the offset. We therefore explore whether calibrating the sensors will make them more difficult to fingerprint. We note that calibration has a side effect of improving the accuracy of sensor readings and is therefore of independent value. We perform the calibration only on the sensors in our 30 lab smartphones because we felt that calibration is too time consuming for the volunteers. Moreover, we could better control the quality of the calibration process when carried out in the lab.

\begin{figure*}[!htb]
\centering
\begin{tabular}{lccc}
\epsfig{file=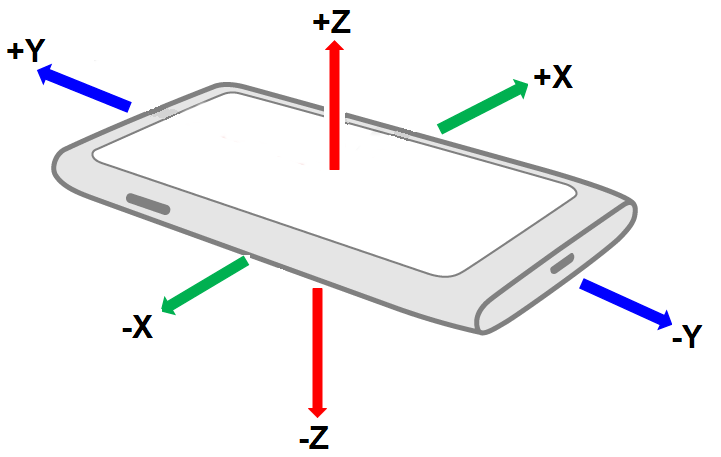,width=0.25\linewidth,clip=}&\epsfig{file=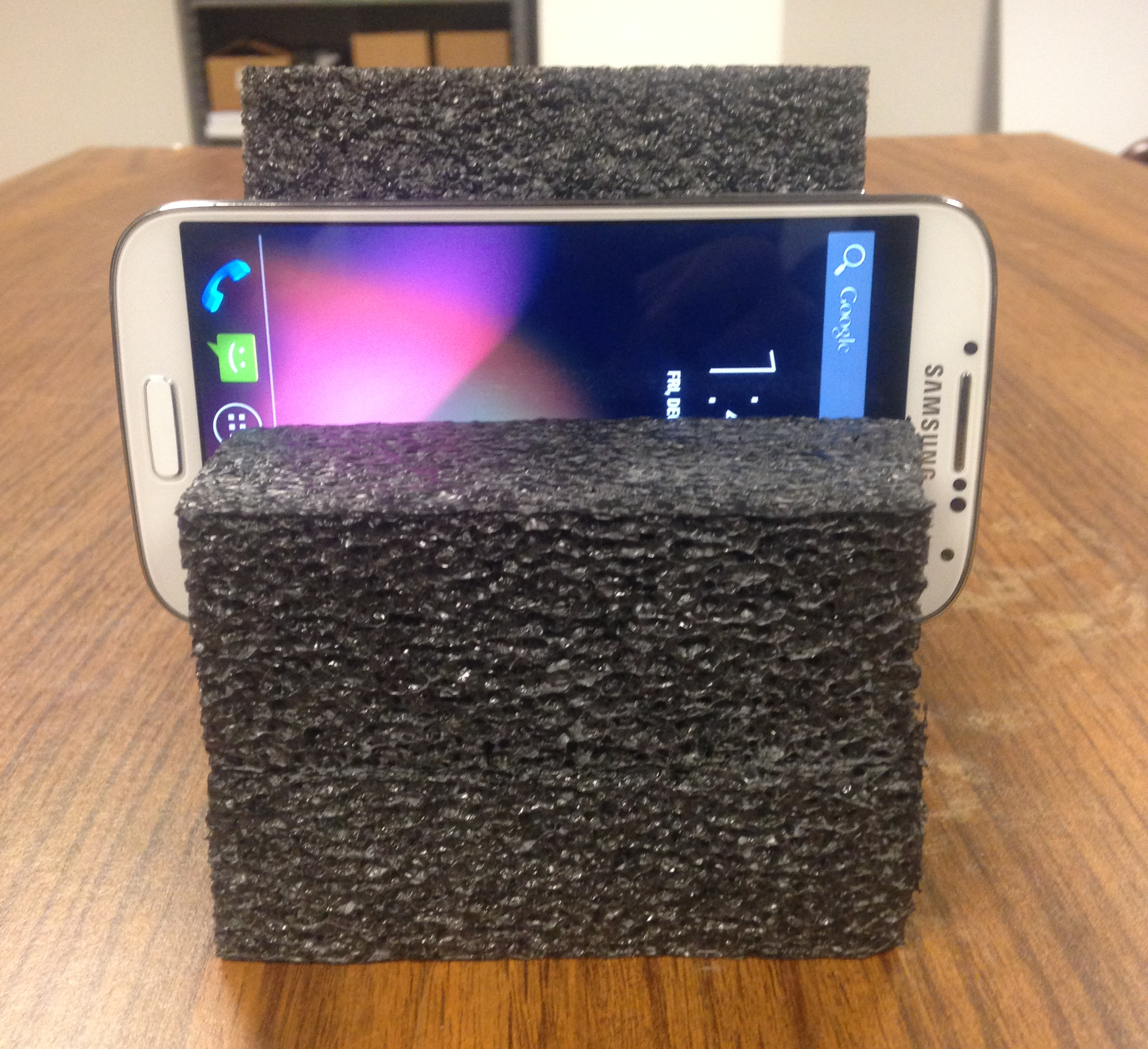,width=0.21\linewidth,height=0.21\linewidth,clip=}& 
\epsfig{file=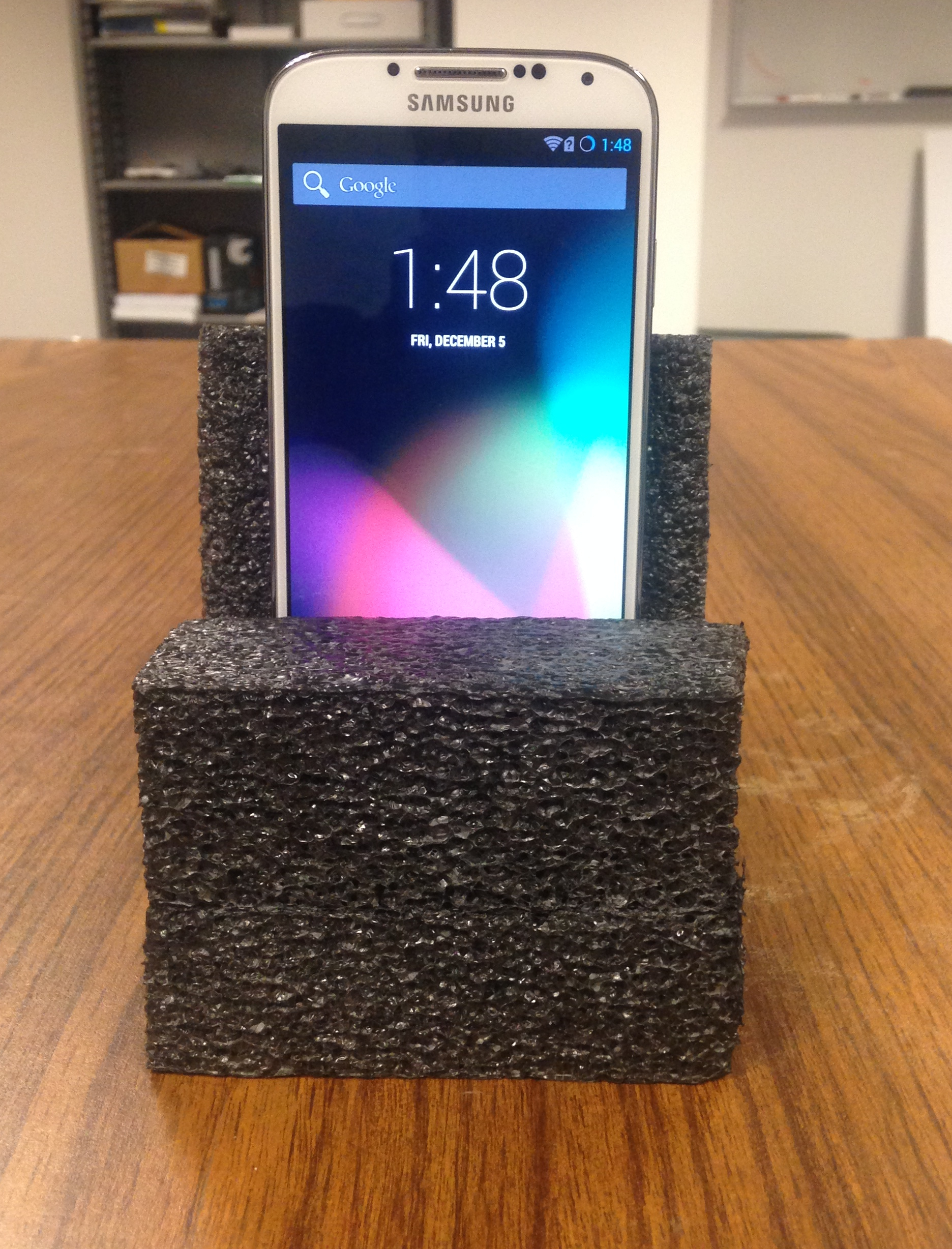,width=0.21\linewidth,height=0.21\linewidth,clip=}&\epsfig{file=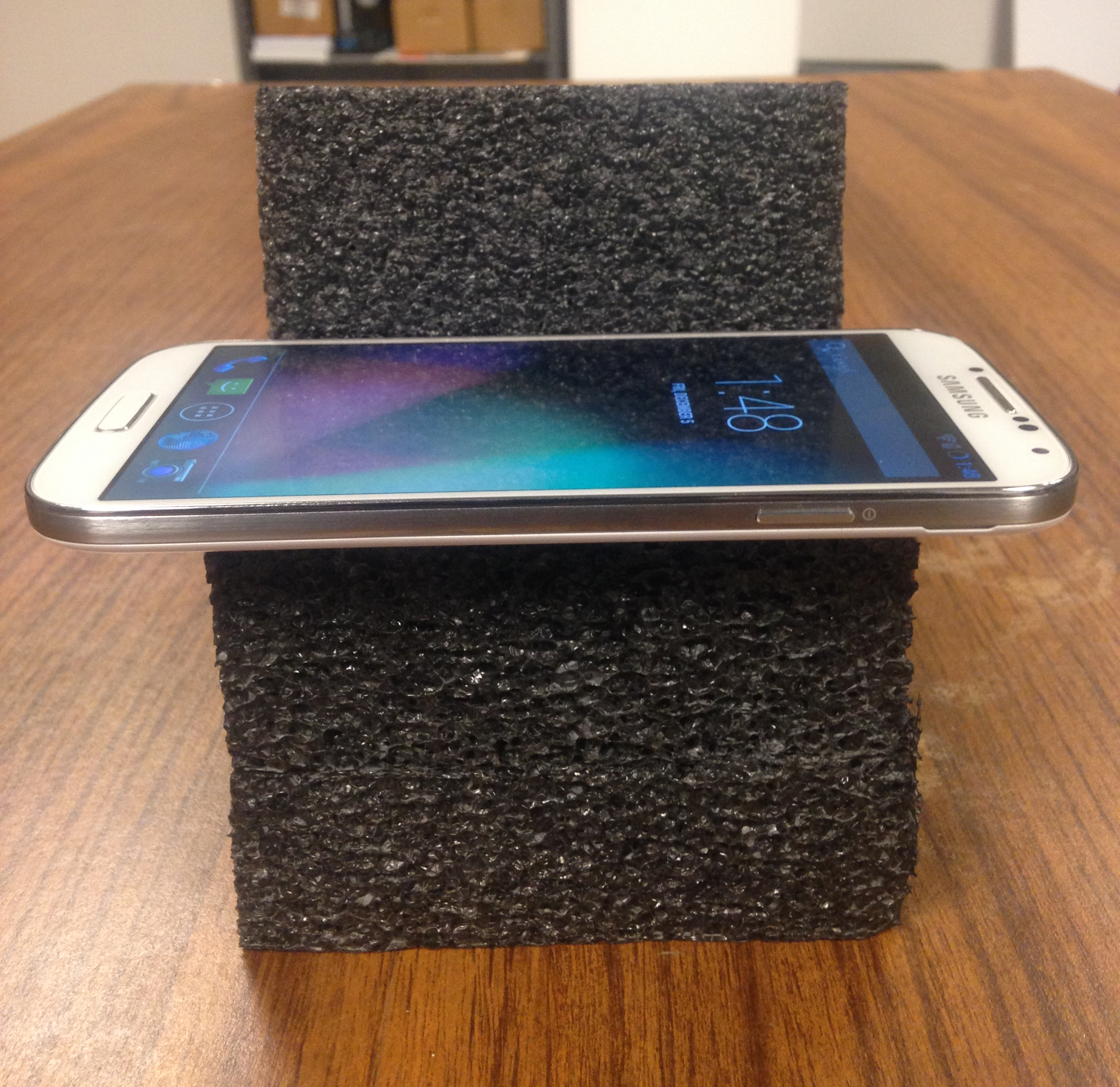,width=0.21\linewidth,height=0.21\linewidth,clip=}
\end{tabular}
\caption{Calibrating accelerometer along three axes. We collect measurements along all 6 directions ($\pm x,\pm y, \pm z$).} 
\label{accelcalibration}
\end{figure*}

First, let us briefly describe the sensor coordinate system as the sensor framework using a standard 3-axis coordinate system to express data 
values. For most sensors, the coordinate system is defined relative to the device's screen when the device is held in its default orientation (shown 
in figure~\ref{accelcalibration}). When the device is held in its default orientation, the positive $x$-axis is horizontal and points to the right, the 
positive $y$-axis is vertical and points up, and the positive $z$-axis points toward the outside of the screen face\footnote{Android and iOS 
consider the positive and negative direction along an axis differently.}. We compute offset and gain error in all three axes.

\begin{figure}[!h]
\centering
\includegraphics[width=0.85\columnwidth]{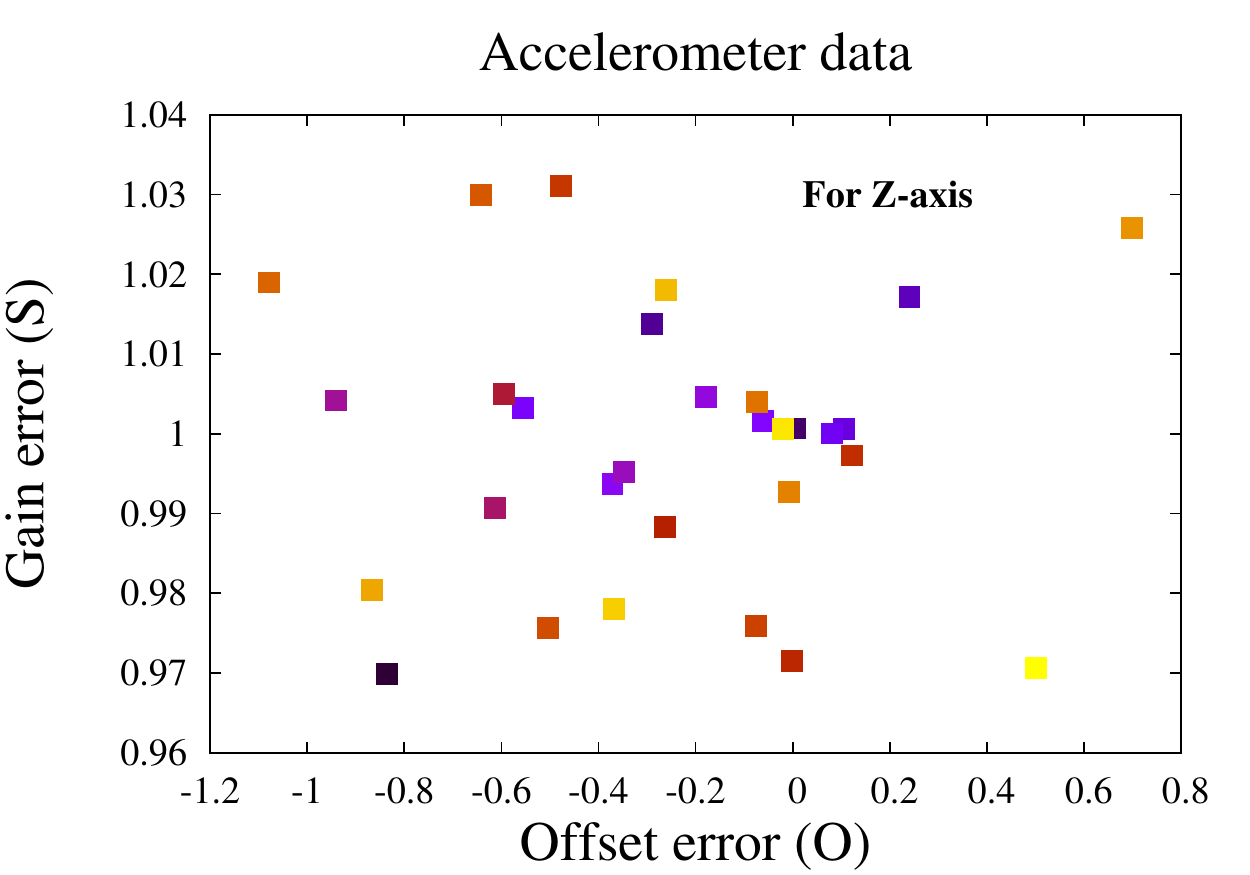}
\caption{Accelerometer offset and gain error of from 30 smartphones.} 
\label{acceloffsetgainplot}
\end{figure}

\paragraphb{Calibrating the Accelerometer:}
Considering both offset and gain error, the measured output of the accelerometer ($a^M=[a_x^M,a_y^M,a_z^M]$) can be expressed as:
\begin{equation}
\left[ \begin{array}{c}
       a_x^M\\
       a_y^M\\
       a_z^M
       \end{array}
\right] = \left[ \begin{array}{c}
       O_x\\
       O_y\\
       O_z
       \end{array}
\right] +  \begin{bmatrix}
       S_x  & 0  &  0\\
       0  &  S_y & 0\\
       0  &  0  &  S_z
       \end{bmatrix}
       \left[ \begin{array}{c}
       a_x\\
       a_y\\
       a_z
       \end{array}
\right] 
\end{equation}
where $S=[S_x, S_y, S_z]$ and $O=[O_x, O_y, O_z]$ respectively represents the gain and offset errors along all three axes ($a=[a_x, a_y, a_z]$ refers 
to the actual acceleration). In the ideal world $[S_x, S_y, S_z]=[1,1,1]$ and $[O_x, O_y, O_z]=[0,0,0]$, but in reality they differ from the desired 
values. To compute the offset and gain error of an axis, we need data along both the positive and negative direction of that axis (one measures 
positive $+g$ while the other measures negative $-g$). In other words, six different \emph{static} positions are used where in each position one of 
the axes is aligned either along or opposite to earth's gravity. This causes the $a=[a_x, a_y, a_z]$ vector to take one of the following six possible 
values $\{[\pm g,0,0], [0,\pm g,0], [0,0,\pm g]\}$. For example, if $a_{z+}^M$ and $a_{z-}^M$ are two values of accelerometer reading along the 
positive and negative $z$-axis, then we can compute the offset ($O_z$) and gain ($S_z$) error using the following equation:\nolinebreak
\begin{equation}
S_z = \frac{a_{z+}^M-a_{z-}^M}{2g} , \hspace{16pt} O_z = \frac{a_{z+}^M-a_{z-}^M}{2} \label{offsetgain}
\end{equation}
We take 10 measurements along all six directions ($\pm x,\pm y, \pm z$) from all our lab devices as shown in Figure~\ref{accelcalibration}. From 
these measurements we compute the average offset and gain error along all three axes using equation (\ref{offsetgain}). 
Figure~\ref{acceloffsetgainplot} shows a scatter-plot of the errors along $\mathtt{z-axis}$ for 30 smartphones (each point represents a 
single device). We can see that the devices are scattered around allover the plot which signifies that different devices have different amount of 
offset and gain error. Such unique distinction makes fingerprinting feasible.

\begin{figure}[!htb]
\centering
\begin{tabular}{c}
\epsfig{file=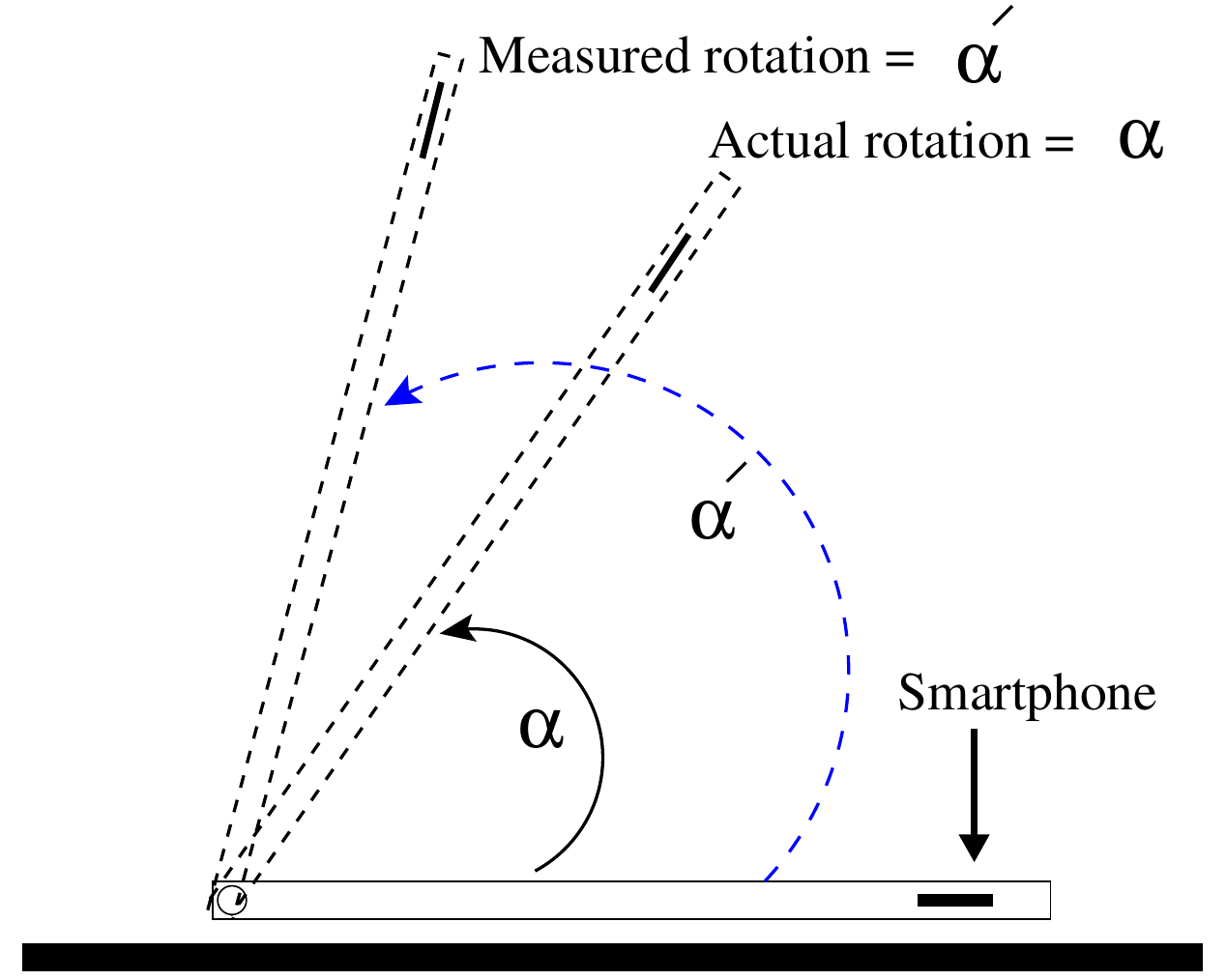,width=0.5\columnwidth,clip=}\\
(a)\\
\epsfig{file=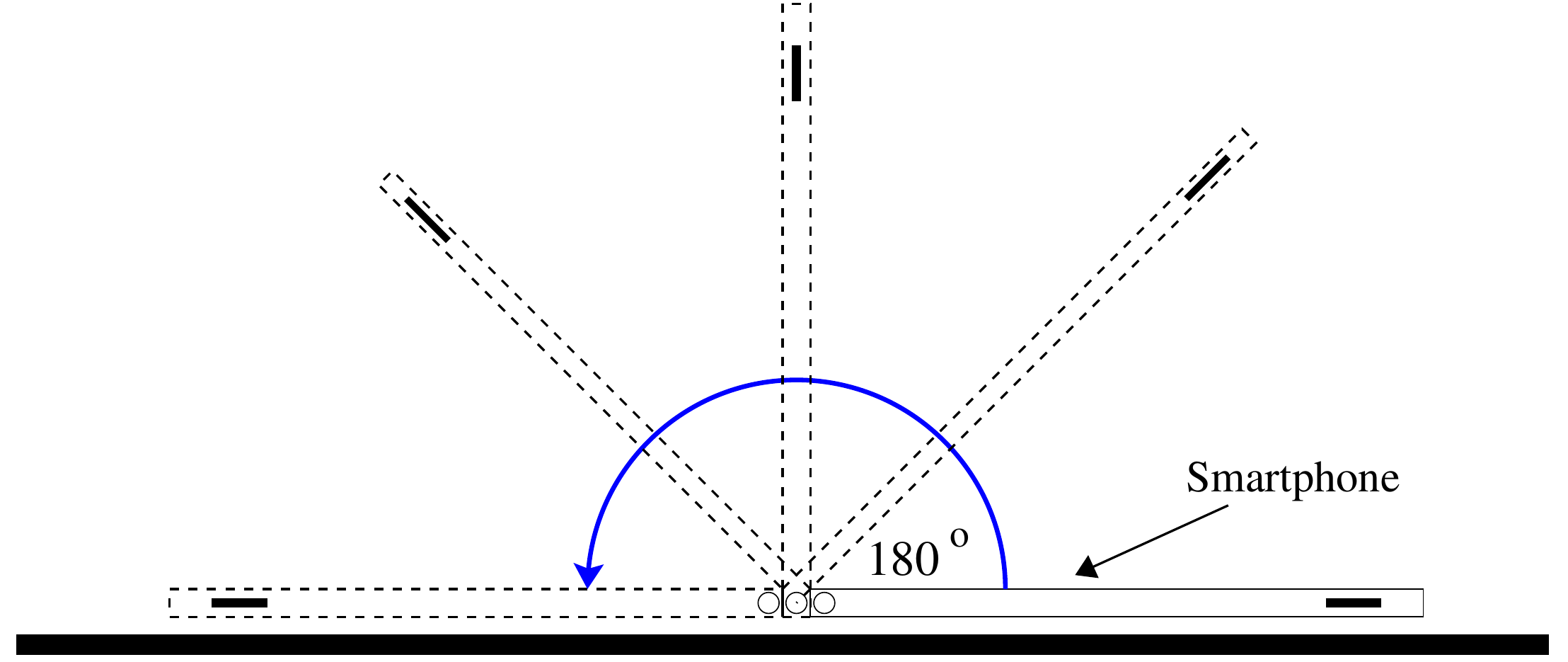,width=0.8\columnwidth,clip=}\\
(b)
\end{tabular}
\caption{a) Offset and gain error in gyroscope impact systems that use them for angular displacement measurements. b) 
Calibrating the gyroscope by rotating the device by $180^\circ$ in the positive $x$-axis direction.} 
\label{gyrocalibration}
\end{figure}

\paragraphb{Calibrating the Gyroscope:}
Calibrating gyroscope is a harder problem as we need to induce a fixed angular change to determine the offset and gain error. Similar to 
accelerometer we can also represent the measured output of the gyroscope ($\omega^M=[\omega_x^M,\omega_y^M,\omega_z^M]$) using the following equation:
\begin{equation}
\left[ \begin{array}{c}
       \omega_x^M\\
       \omega_y^M\\
       \omega_z^M
       \end{array}
\right] = \left[ \begin{array}{c}
       O_x\\
       O_y\\
       O_z
       \end{array}
\right] + \begin{bmatrix}
       S_x  & 0  &  0\\
       0  &  S_y & 0\\
       0  &  0  &  S_z
       \end{bmatrix} 
       \left[ \begin{array}{c}
       \omega_x\\
       \omega_y\\
       \omega_z
       \end{array}
\right]  
\end{equation}
where again $S=[S_x, S_y, S_z]$ and $O=[O_x, O_y, O_z]$ respectively represents the gain and offset errors along all three axes. 
Here, $\omega=[\omega_x,\omega_y,\omega_z]$) represents the ideal/actual angular velocity. Ideally all gain and offset errors should be equal to $1$ 
and $0$ respectively. But in the real world when the device is rotated by a fixed amount of angle, the measured angle tends to deviate from the 
actual angular displacement (shown in figure~\ref{gyrocalibration}(a)). This impacts any system that uses gyroscope for angular-displacement 
measurements.

To calibrate gyroscope we again need to collect data along all six different directions ($\pm x,\pm y,\pm z$) individually, but this time 
instead of keeping the device stationary we need to \emph{rotate} the device by a fixed amount of angle ($\theta$). In our setting, we set 
$\theta={180}^{\circ}$ (or $\pi$ $\mathtt{rad}$). For example, Figure~\ref{gyrocalibration}(b) shows how we rotate the the smartphone by 
${180}^{\circ}$ around the positive $x$-axis. The angular displacement along any direction can be computed from gyroscope data in the 
following manner:
\begin{align}
\omega_i^M &= O_i + S_i\omega, \hspace{8pt}  \mbox{$i\in\{\pm x,\pm y,\pm z\}$}\nonumber \\
\int_0^{t}\omega_i^M\,\mathrm{d}t &= \int_0^{t}O_i\,\mathrm{d}t + S_i \int_0^{t}\omega\,\mathrm{d}t\nonumber\\
\theta_i^M & = O_i {t} + S_i \theta
\label{calibrategyroeqn}
\end{align}
where $t$ refers to the time it took to rotate the device by $\theta$ angle with a fixed angular velocity of $\omega$. Now, for any two measurements 
along the opposite directions of an axis we can compute the offset and gain error using the following equation:\nolinebreak
\begin{equation}
O_i = \frac{\theta_{i+}^M+\theta_{i-}^M}{t_1-t_2}, \hspace{4pt} S_i = \frac{\theta_{i+}^M - \theta_{i-}^M - O_i(t_1-t_2)}{2\pi}
\end{equation}
where $i\in\{x,y,z\}$ and $t_1$ and $t_2$ represents the timespan of the positive and negative measurement respectively. We take 10 measurements along 
all six directions ($\pm x,\pm y, \pm z$) and compute the average offset and gain error along all three axes. However, since its practically 
impossible to manually rotate the device a fixed angular velocity, the integration in equation (\ref{calibrategyroeqn}) will introduce noise and 
therefore, the calculated errors will at best be approximations of the real errors. We also approximate the integral using trapezoidal rule which 
will introduce some more errors. 

\begin{figure}[!h]
\centering
\includegraphics[width=0.85\columnwidth]{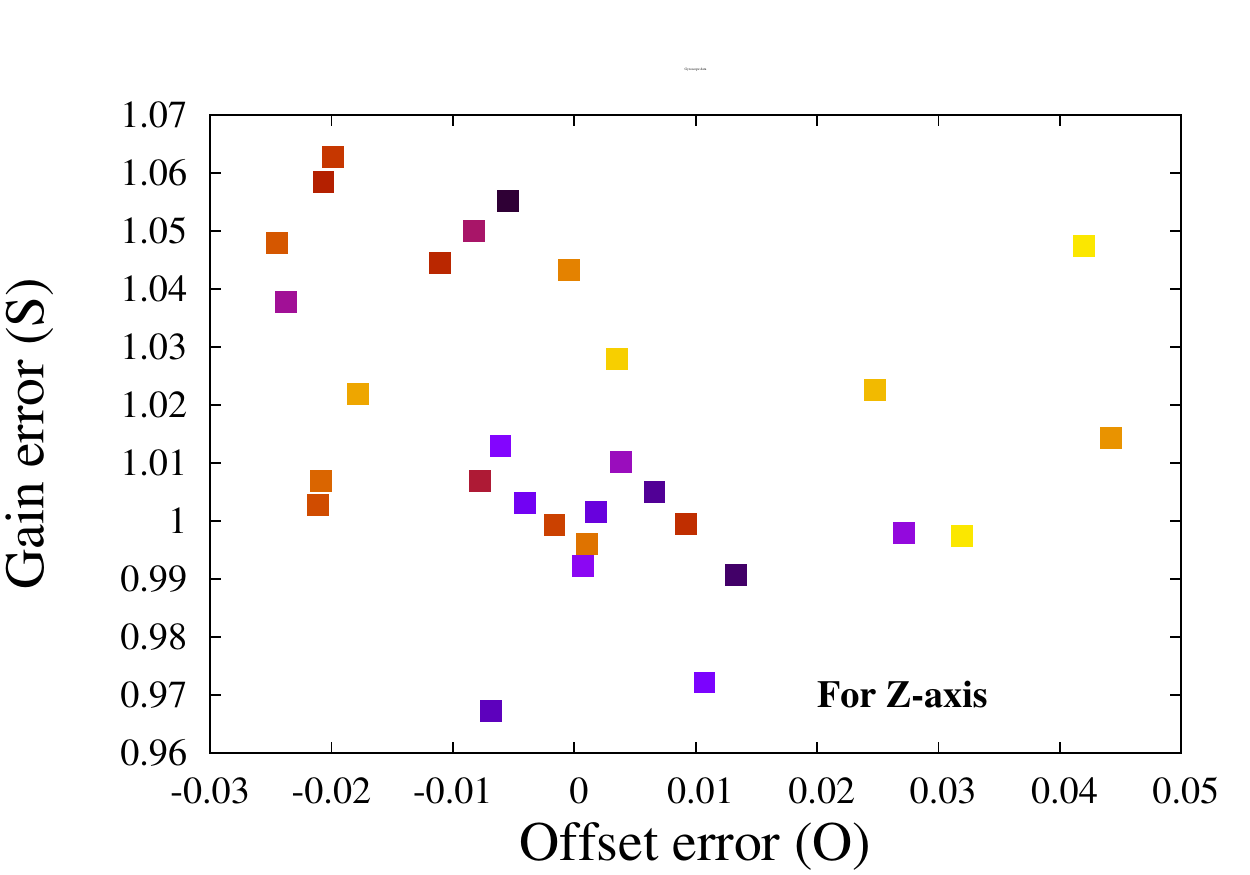}
\caption{Gyroscope offset and gain error of 30 smartphones.} 
\label{gyrooffsetgainplot}
\end{figure}

We next visualize the offset and gain error obtained from the gyroscopes of 30 smartphones (only showing for $z$-axis). 
Figure~\ref{gyrooffsetgainplot} shows our findings. We see similar results compared to accelerometers where devices are scattered around at different 
regions of the plot. This suggests that gyroscopes exhibit different range of offset of and gain error across different units.  

\begin{figure}[!htb]
\centering
\begin{tabular}{c}
\epsfig{file=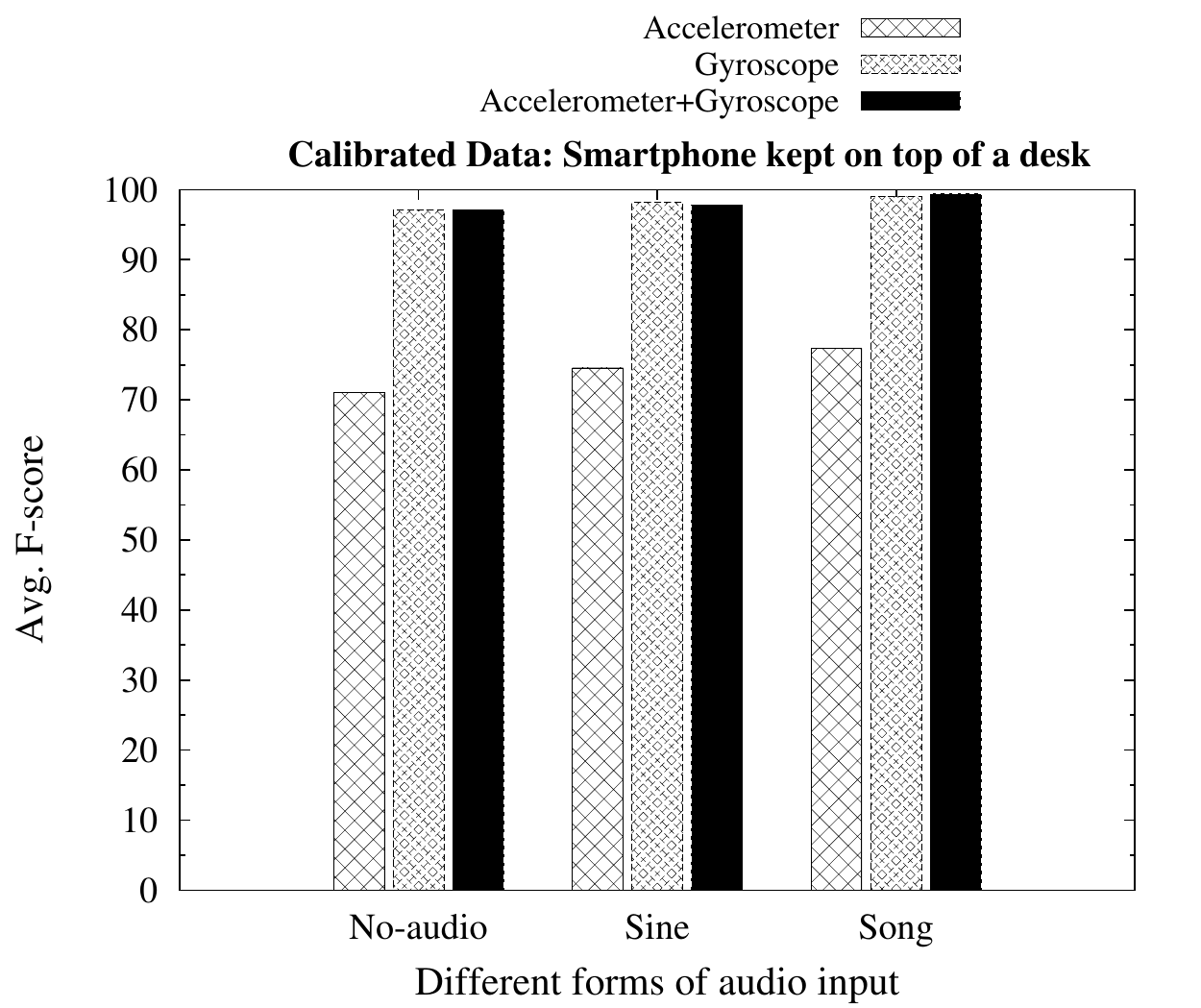,width=0.85\columnwidth,clip=}\\
(a)\\
\epsfig{file=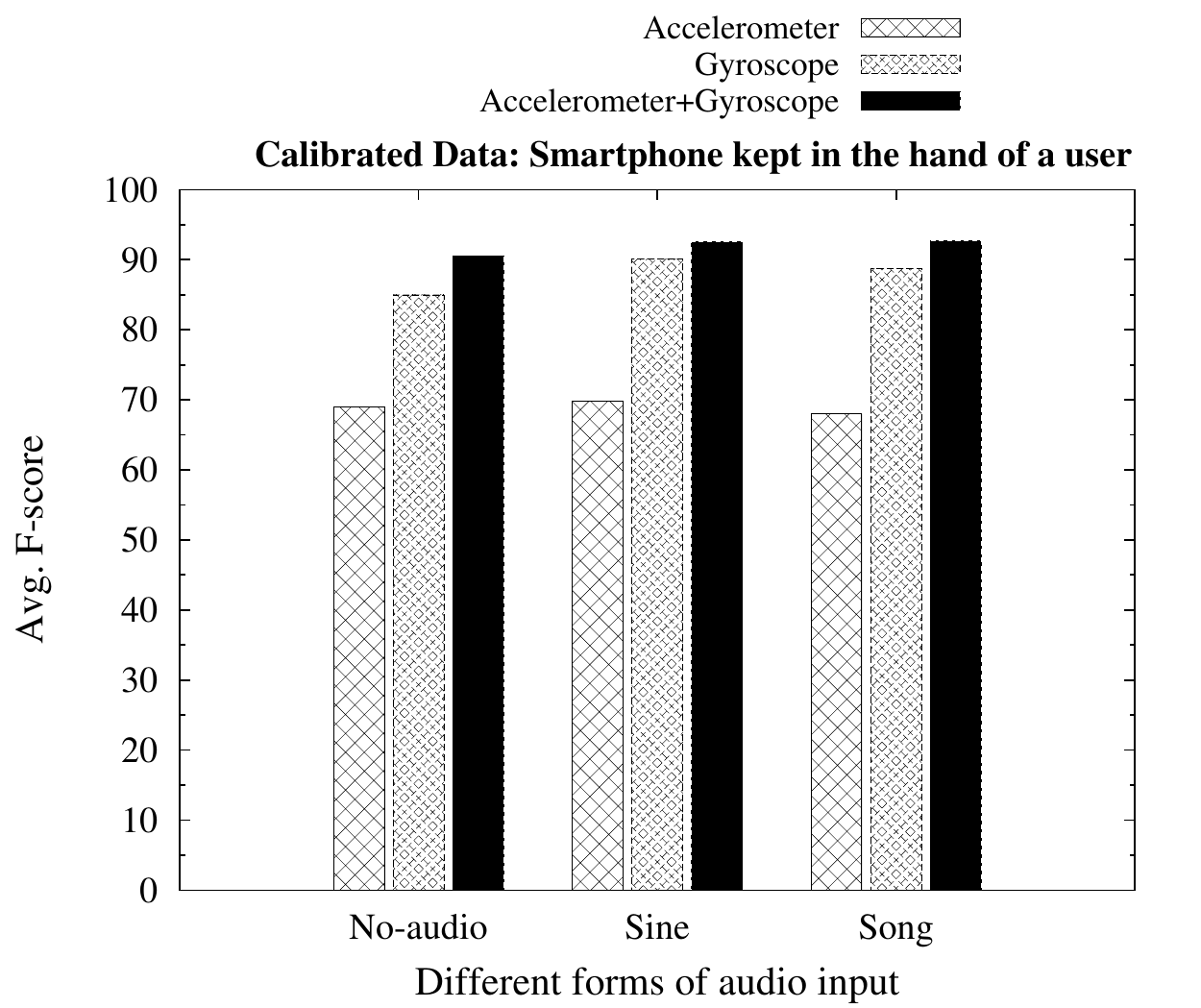,width=0.85\columnwidth,clip=}\\
(b)
\end{tabular}
\caption{Average F-score for calibrated data under lab setting. Results obtained from 30 smartphones where the smartphones were kept a) \emph{on top 
of a desk} b) \emph{in the hand} of the user while collecting sensor data.} 
\label{lab_ondesk_inhand_calibrated}
\end{figure}

\paragraphb{Fingerprinting Calibrated Data:}
In this section we look at how calibrating sensors impacts the fingerprinting accuracy. For this setting, we first correct the raw values by removing the 
the offset and gain errors before extracting features from them. That is, the calibrated value $a^C = a^M/g - o$. We then generate fingerprints on the corrected data and 
train the classifiers on the new fingerprints. Figure~\ref{lab_ondesk_inhand_calibrated} shows the average F-score for calibrated data under three 
scenarios, considering both cases where the devices were kept on top of desk and in the hand of a user. When we compare the results from uncalibrated 
data (figure~\ref{lab_ondesk_inhand}) to those from calibrated data, we see that the F-score reduces by almost 30\% for accelerometer data but not as 
much for the gyroscope data. This suggests that we were able to calibrate the accelerometer much more precisely than the gyroscope, as expected given the more complex and error-prone manual calibration procedure for the gyroscope. Another interesting observation is that audio stimulation provides a significant improvement in classifier accuracy. This suggests that audio stimulation does not influence (and perhaps even hinders) the dominant features removed by the calibration, but does significantly impact secondary features that come into play once calibration is carried out.
Overall, our results demonstrate that calibration is a promising technique, especially if more precise measurements can be made. Manufacturers should be encouraged to perform better calibration to both improve the accuracy of their sensors and to help protect users' privacy.

\subsection{Data Obfuscation}
\paragraphb{Basic Obfuscation:}
Rather than remove the calibration errors, we can instead add extra noise to hide the calibration. This approach has the advantage of not requiring a calibration step, which requires user intervention and is particularly difficult for the gyroscope sensors. As such, the obfuscation technique could be deployed with an operating system update. Obfuscation, however, adds extra noise and can therefore negatively impact the utility of the sensors (in contrast to calibration, which improves their utility). We therefore first consider small obfuscation values in the range that is similar to what we observed in the calibration errors above. Adding  noise in this range is roughly equivalent to switching to a differently (mis)calibrated phone and therefore should cause minimal impact to the user.

\begin{figure}[!htb]
\centering
\begin{tabular}{cc}
\epsfig{file=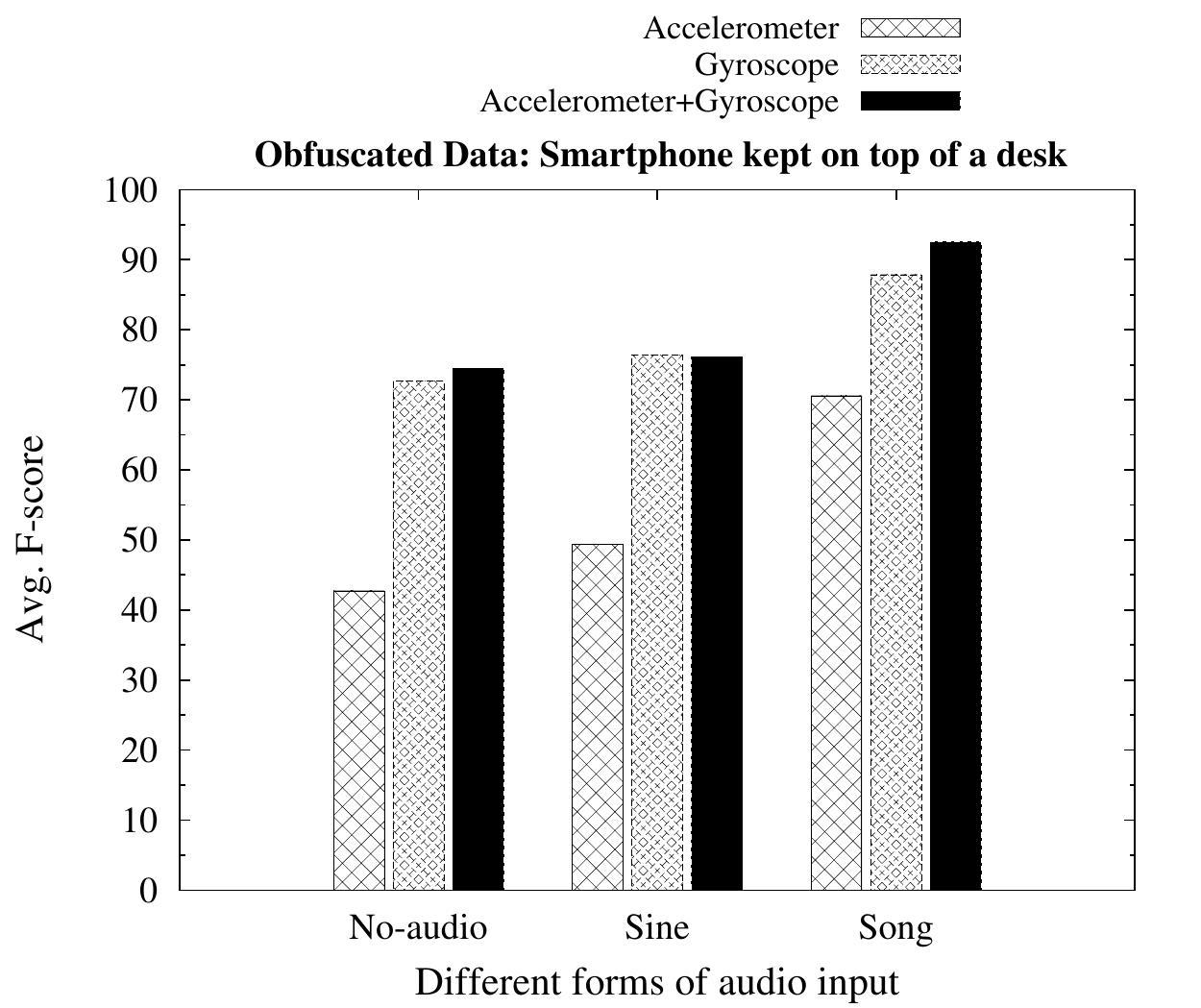,width=0.85\columnwidth,clip=}\\
(a)\\
\epsfig{file=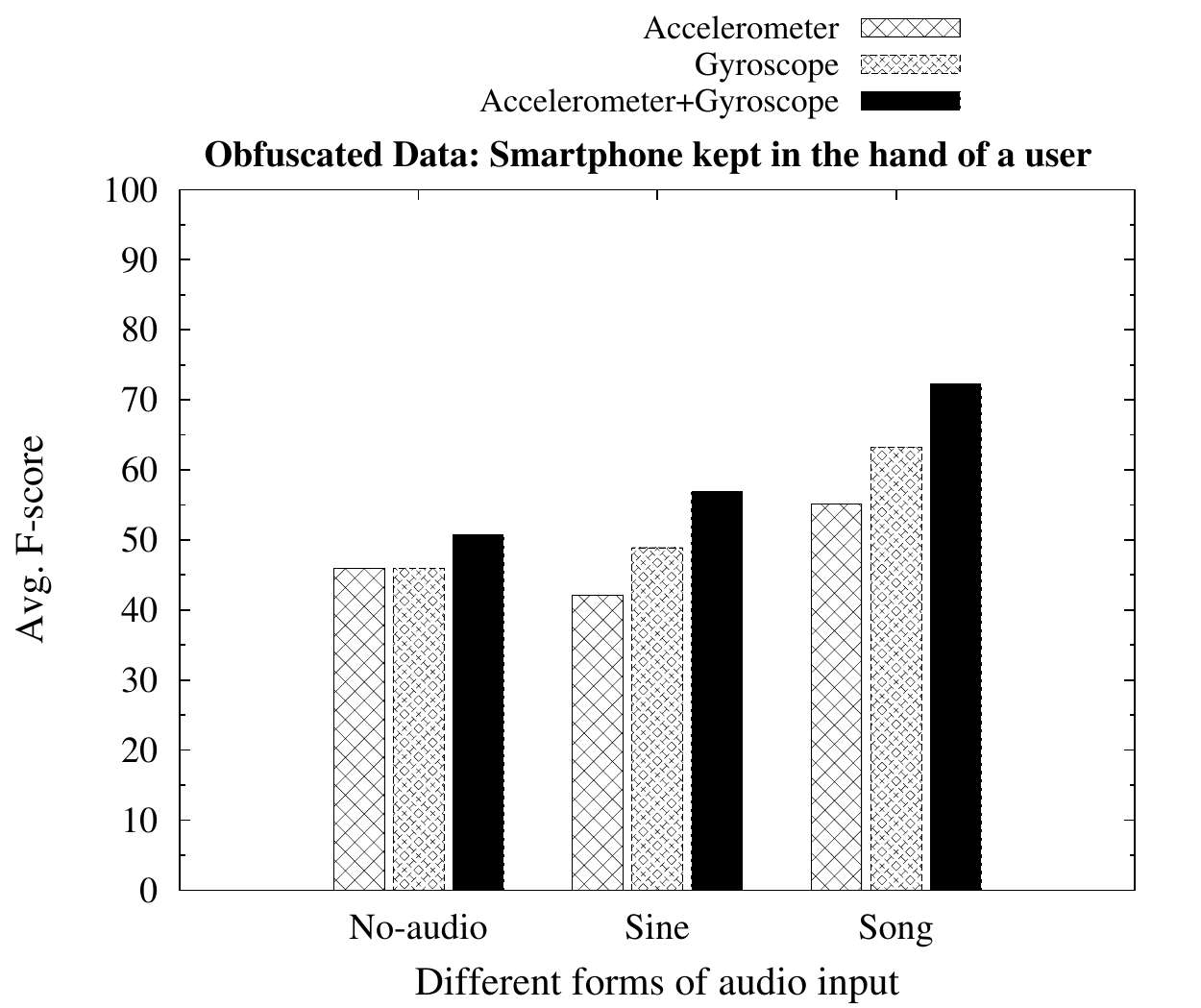,width=0.85\columnwidth,clip=}\\
(b)
\end{tabular}
\caption{Average F-score for obfuscated data under lab setting. Results obtained from 30 smartphones where 
the smartphones were kept a) \emph{on top of a desk} b) \emph{in the hand} of the user while collecting sensor data.} 
\label{lab_ondesk_inhand_obfuscation}
\end{figure}

To add obfuscation noise, we compute $a^O = a^M*g^O+o^O$, where $g^O$ and $o^O$ are the obfuscation gain and offset, respectively. Based on Figures~\ref{acceloffsetgainplot} and \ref{gyrooffsetgainplot}, we choose a range of [-0.5,0.5] for the accelerometer offset, [-0.1,0.1] for the gyroscope offset, and [0.95,1.05] for the gain. For each session, we pick uniformly random obfuscation gain and offset values from the range; by varying the obfuscation values we make it difficult to fingerprint repeated visits. Figure~\ref{lab_ondesk_inhand_obfuscation} 
summarizes our findings when we apply obfuscation to all the sensor data obtained from our 30 lab smartphones. Compared to unaltered data 
(figure~\ref{lab_ondesk_inhand}), data obfuscation seems to provide significant improvement in terms of reducing the average F-score. Depending on the 
type of audio stimulation F-score reduces by almost 10--25\% when smartphones are kept stationary on the desk and by 20--45\% when smartphones are 
kept stationary in the hand of the user. The 
impact of audio stimulation in fingerprinting motion sensors is much more visible in these results. We see that F-score increases by almost 15\% when 
a song is being played in the background; again, we expect this to be a consequence of us having hidden the calibration errors that are the primary discriminant between phones.

\begin{figure}[!htb]
\centering
\begin{tabular}{cc}
\epsfig{file=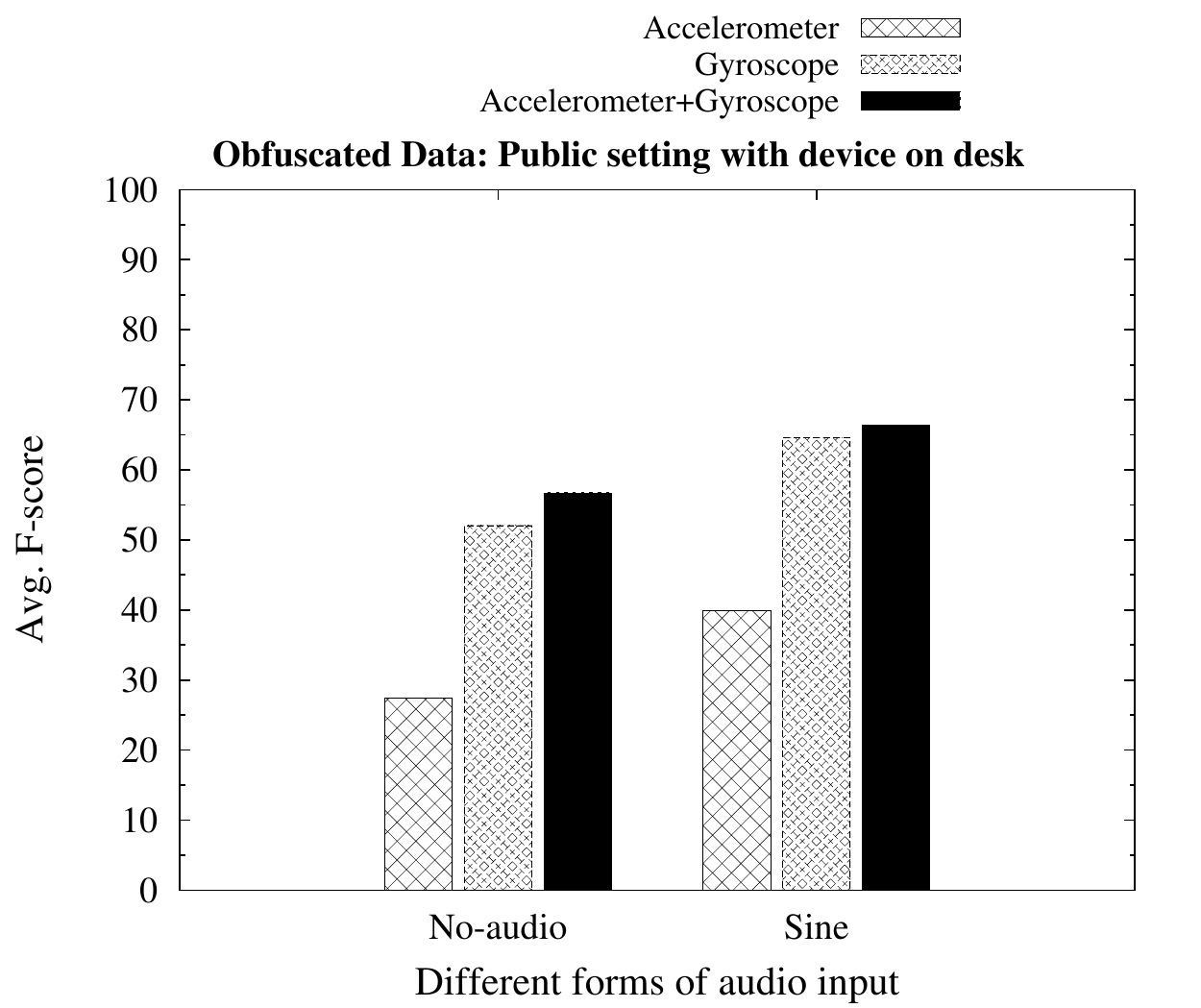,width=0.85\columnwidth,clip=}\\
(a)\\
\epsfig{file=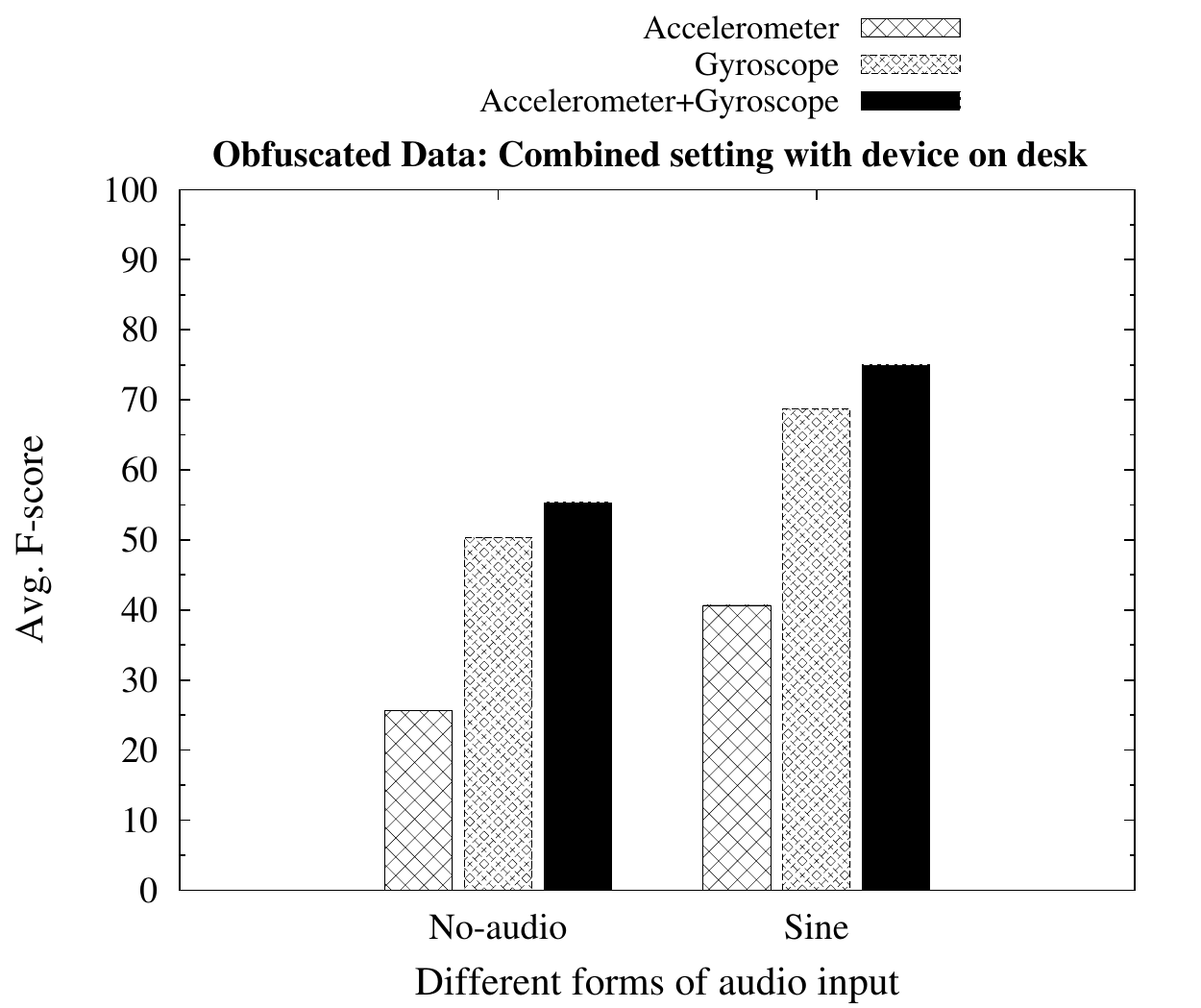,width=0.85\columnwidth,clip=}\\
(b)
\end{tabular}
\caption{Average F-score for different forms of audio stimulation for obfuscated data. Results obtained from  a) 63 public smartphones b) 93 
smartphones (by combining the 63 public smartphones with our 30 lab phones) where the smartphones were kept \emph{on top of a desk} while collecting 
sensor data.} 
\label{public_combo_obfuscation}
\end{figure}

Next, we apply similar techniques to the public and combined dataset. We apply the same range of offset and gain errors to the raw values before 
generating fingerprints. Figure~\ref{public_combo_obfuscation} summarizes our results for both presence and absence of audio stimulation. We see that 
F-score reduces by approximately 20--40\% (Figure~\ref{lab_ondesk_inhand_calibrated}(a)). We expect the lower accuracy is a consequence of a larger data set, suggesting that for even larger sets the impact of obfuscation is likely to be even more pronounced.

\paragraphb{Increasing the Obfuscation Range:}
We next look at 
how the fingerprinting technique reacts to different ranges of obfuscation. Starting with our base ranges of $[-0.5,0.5]$ and $[-0.1,0.1]$ for the accelerometer and gyroscope offsets, respectively, and $[0.95,1.05]$ for the gain, we linearly scale the ranges and observe the impact on the average F-score. We scale all ranges by the same amount, increasing the ranges symmetrically on both sides of the interval midpoint.

\begin{figure}[!htb]
\centering
\includegraphics[width=0.85\columnwidth]{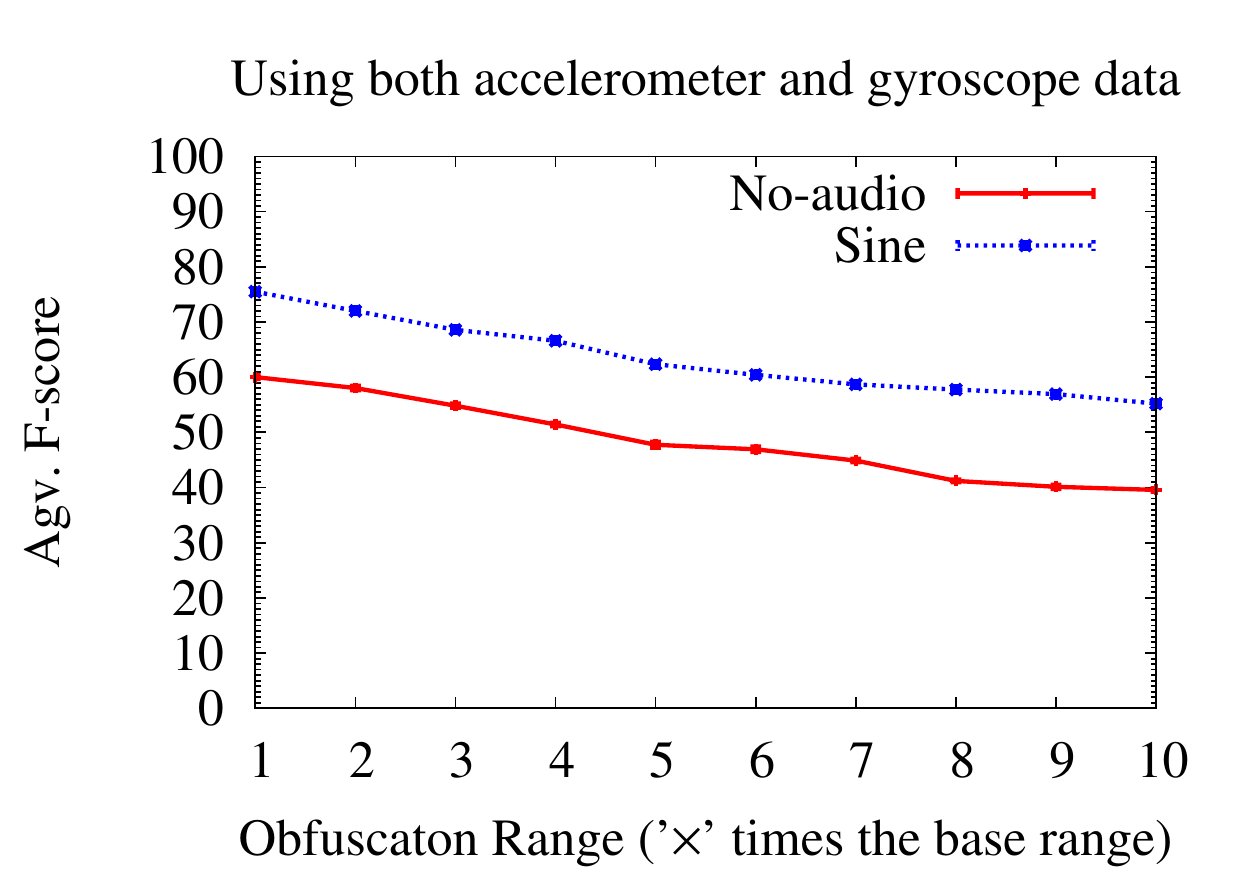}
\caption{Impact of obfuscation range as the range is linearly scaled up from  $1\mathtt{x}$ to  $10\mathtt{x}$ of the base range.}
\label{obfuscationrange}
\end{figure}

For this experimental setup we only consider the combined dataset as this contains the most number of devices (93 in total). We also restrict 
ourselves to the setting where we combine both the accelerometer and gyroscope features because this provides the optimal result (as evident from 
all our past results). Figure~\ref{obfuscationrange} highlights our findings. As we can see increasing the obfuscation range does reduce F-score but 
it has a \emph{diminishing return}. For 10x increment, the F-score drops down to approximately 40\% and 55\% for no-audio and audio stimulation 
respectively. Beyond $10\mathtt{x}$ increment (not shown) the reduction in F-score is minimal (at most 10\% reduction at $50\mathtt{x}$ increment). This result 
suggests that simply obfuscating the raw values is not sufficient to hide all unique characteristics of the sensors. So far we have only manipulated 
the signal value but did not alter any of the frequency features and as a result the classifier is still able to utilize the spectral features to 
uniquely distinguish individual devices.

\paragraphb{Enhanced Obfuscation:}
Given that we know that the spectral features are not impacted my our obfuscation techniques, we now focus on adding noise to the frequency of the 
sensor signal. Our data injection procedure is described in algorithm~\ref{datainjection}. The main idea is to probabilistically insert a modified 
version of the current data point in between the past and current timestamp where the timestamp itself is randomly selected. Doing so will influence 
cubic interpolation of the data stream which in turn will impact the spectral features extracted from the data stream.

\begin{algorithm}[h]
\caption{Obfuscated Data Injection}\label{datainjection}
\begin{algorithmic}
\STATE  {\bf Input:} {Time series Data $D[t]$, Probability $Pr$,\\\hspace{20pt}Obfuscation Range $\mathit{Obf}_{range}$, Offset $O$, Gain $S$}
\STATE {\bf 
Output:} {Modified Data Stream $MD[t]$}
\STATE{$last_{timestamp}\leftarrow Null$}
\STATE{$\mathit{offset}\leftarrow Null$}
\STATE{$gain\leftarrow Null$}
\STATE {\#Random(range) : randomly selects a value in range}
\FOR{$i = 1$ to $length(D)$}
    \STATE {\underline{\#New data insertion}}
    \IF{$i>1$ and $Random([0,1])<Pr$ }
      \STATE {$\mathit{offset}\leftarrow Random(\mathit{Obf}_{range})$}
      \STATE {$gain\leftarrow Random(\mathit{Obf}_{range})$}
      \STATE {$time\leftarrow Random([i,last_{timestamp}])$}
      \STATE {$D[time]\leftarrow InsertData(D[i],\mathit{offset},gain)$}
    \ENDIF
    \STATE {\underline{\#Original Data}}
    \STATE {$D[i]\leftarrow InsertData(D[i],O,S)$}
    \STATE {$last_{timestamp}\leftarrow i$ }
\ENDFOR
\STATE {\textbf{return} $MD$}
\end{algorithmic}
\end{algorithm}

\begin{figure}[!htb]
\centering
\includegraphics[width=0.85\columnwidth]{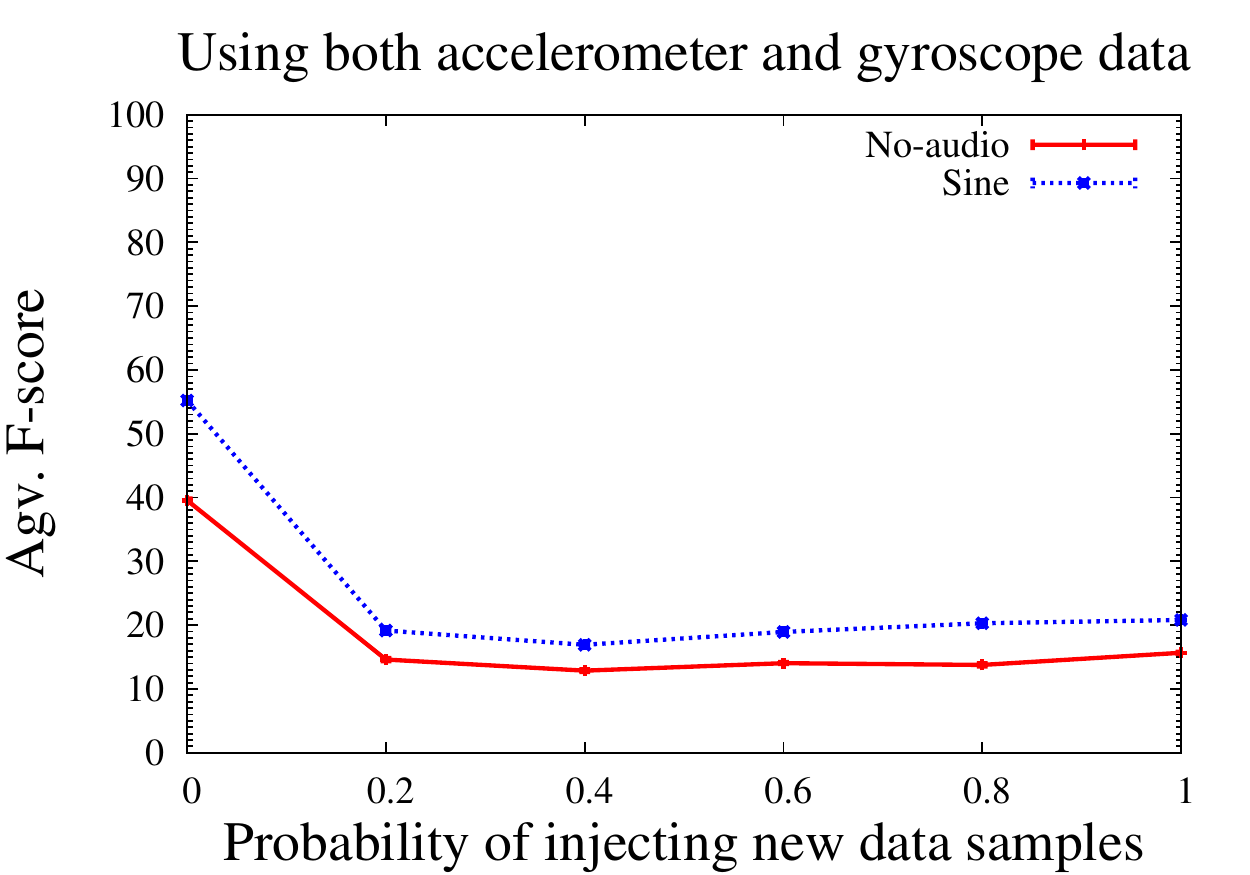}
\caption{Impact of randomly inserting new data points.}
\label{obfuscationradomization}
\end{figure}

To evaluate our approach we first fix a obfuscation range. We choose $10\mathtt{x}$ of the base range from the previous section as our fixed 
obfuscation range. We then vary the probability of data injection from [0,1]. Figure~\ref{obfuscationradomization} shows our findings. We can see 
that even with relatively small amount of data injection ($\leq 0.4$) we can reduce the average F-score to $\approx$15--20\% depending on what type 
of input stimulation is applied.

%% file: limitation.tex
\section{Deployment Considerations}{\label{deployment}}
We envision our obfuscation technique as a middle-ware, sitting between the OS and user application. Under default setting data is always obfuscated 
unless the user explicitly allows an application to access unaltered sensor data. For example, a 3-D game might need access to raw accelerometer and 
gyroscope data instead of the obfuscated data to operate properly, in which case this will be noticeable to the user who can then provide 
the appropriate permission to the application. Our default obfuscated setting will ensure that users do not have to worry about applications like 
browser accessing sensor data without their awareness.

%% file: conclusion.tex
\section{Conclusion}{\label{conclusion}}
In this paper we show that motion sensors such as accelerometers and gyroscopes can be used to uniquely identify smartphones. The more concerning 
matter is that these sensors can be surreptitiously accessed from the browser without user awareness. We also show that injecting audio stimulation 
in the background improves detection rate as sensors like gyroscopes react to acoustic stimulation differently. 

Our countermeasure techniques, however, mitigate such threats by obfuscating anomalies in sensor data. We were able to significantly reduce 
fingerprinting accuracy by employing a simple, yet effective obfuscation technique that injects random data points inside the generated  
sensor data-stream. As a general conclusion we suggest using our obfuscation technique in the absence of explicit user permission/awareness.

%% file: appendix.tex
\section{Feature Description}{\label{appendix_features}}

\paragraphb{Mean Signal Value:} 
This feature computes the arithmetic mean of a signal amplitude. In the case of a set of $N$ values $\{x_1,x_2,\dots,x_N\}$, the mean value is given 
by the following formula:\nolinebreak
\begin{equation}
\mu = \frac{1}{N} \left( x_1 + x_2 +\cdots + x_N \right)
\end{equation}
The mean value provides an approximation of the average signal strength.

\paragraphb{Signal Variance:} 
This feature computes the dispersion in signal strength. For a set of $N$ values $\{x_1,x_2,\dots,x_N\}$, the standard deviation is given 
by the following formula:\nolinebreak
\begin{equation}
\sigma = \sqrt{\frac{1}{N} \sum_{i=1}^{N}(x_i-\mu)^2}
\end{equation}
where $\mu$ refers to the mean signal strength. Variance measures the spread of a signal strength.

\paragraphb{Average Deviation:} 
This feature measures the average distance from mean. In the case of a set of $N$ values $\{x_1,x_2,\dots,x_N\}$, the average deviation is 
computes using the following formula:\nolinebreak
\begin{equation}
AvgDev = \frac{1}{N}  \sum_{i=1}^{N}|x_i-\mu| 
\end{equation}
where $\mu$ refers to the mean signal strength. 

\paragraphb{Skewness:} 
This feature measures asymmetry about mean. For a set of $N$ values $\{x_1,x_2,\dots,x_N\}$, the skewness is computed as:\nolinebreak
\begin{equation}
\gamma_1 = \frac{1}{N} \left( \sum_{i=1}^{N}(\frac{x_i-\mu}{\sigma})^3 \right)
\end{equation}
where $\mu$ and $\sigma$ respectively represents the mean and standard deviation of signal strength. 

\paragraphb{Kurtosis:} 
This feature measures the flatness or spikiness of a distribution. For a set of $N$ values $\{x_1,x_2,\dots,x_N\}$, the kurtosis is computed 
as:\nolinebreak
\begin{equation}
\beta_1 = \frac{1}{N} \left( \sum_{i=1}^{N}(\frac{x_i-\mu}{\sigma})^4 \right)
\end{equation}
where $\mu$ and $\sigma$ respectively represents the mean and standard deviation of signal strength.

\paragraphb{Root-Mean-Square (RMS) Energy:}
This feature computes the square root of the arithmetic mean of the squares of the original audio signal strength at various frequencies. In the case
of a set of $N$ values $\{x_1,x_2,\dots,x_N\}$, the RMS value is given by the following formula:\nolinebreak
\begin{equation}
x_{\mathrm{rms}} = \sqrt{\frac{1}{n} \left( x_1^2 + x_2^2 +\cdots + x_N^2 \right)}
\end{equation}
The RMS value provides an approximation of the average audio signal strength.

\paragraphb{Zero Crossing Rate (ZCR):}
The zero-crossing rate is the rate at which the signal changes sign from positive to negative or back~\cite{chen1988signal}. ZCR for a
signal $s$ of length $T$ can be defined as:\nolinebreak
\begin{equation}
 ZCR = \frac{1}{T}\sum_{t=1}^{T}|s(t)-s(t-1)|
\end{equation}
where $s(t)=1$ if the signal has a positive amplitude at time $t$ and 0 otherwise. Zero-crossing rates provide a measure of the noisiness of the
signal.

\paragraphb{Low Energy Rate:}
The low energy rate computes the percentage of frames (typically 50ms chunks) with RMS power less than the average RMS power for the whole signal.

\paragraphb{Spectral Centroid:}
The spectral centroid represents the ``center of mass'' of a spectral power distribution. It is calculated as the weighted mean of the frequencies
present in the signal, determined using a fourier transform, with their magnitudes as the weights:\nolinebreak
\begin{equation}
Centroid, \mu= \frac{\sum_{i=1}^{N} f_i \cdot m_i } {\sum_{i=1}^{N}m_i }\label{centroid}
\end{equation}
where $m_i$ represents the magnitude of bin number $i$, and $f_i$ represents the center frequency of that bin.

\paragraphb{Spectral Entropy:}
Spectral entropy captures the spikiness of a spectral distribution. To compute spectral entropy, a Digital Fourier Transform (DFT) of the signal is 
first carried out. Next, the frequency spectrum is converted into a probability mass function (PMF) by normalizing the spectrum using the following 
equation:\nolinebreak
\begin{equation}
w_i=\frac{m_i}{\sum_{i=1}^N m_i}\label{pmf}
\end{equation}
where $m_i$ represents the energy/magnitude of the $i$-th frequency component of the spectrum. $w=(w_1,w_2,\dots,w_N)$
is the PMF of the spectrum and N is the number of points in the spectrum. This PMF can then be used to compute the
spectral entropy using the following equation:\nolinebreak
\begin{equation}
H=\sum_{i=1}^N w_i \cdot log_{2}w_i
\end{equation}
The central idea of using entropy as a feature is to capture the peaks of the spectrum and their location.

\paragraphb{Spectral Spread:}
Spectral spread defines the dispersion of the spectrum around its centroid, i.e., it measures the standard deviation of the spectral distribution. So
it can be computed as:\nolinebreak
\begin{equation}
Spread, \sigma = \sqrt{\sum_{i=1}^{N}\left[(f_i-\mu)^2\cdot w_i\right]}\label{spread}
\end{equation}
where $w_i$ represents the weight of the $i$-th frequency component obtained from equation (\ref{pmf}) and $\mu$
represents the centroid of the spectrum obtained from equation (\ref{centroid}).

\paragraphb{Spectral Skewness:}
Spectral skewness computes the coefficient of skewness of a spectrum. Skewness (third central moment) measures the symmetry of the distribution. A
distribution can be positively skewed in which case it has a long tail to the right while a negatively-skewed distribution has a longer tail to the
left. A symmetrical distribution has a skewness of zero. The coefficient of skewness is the ratio of the skewness to the standard deviation raised to
the third power.\nolinebreak
\begin{equation}
Skewness = \frac{\sum_{i=1}^{N}\left[(f_i-\mu)^3 \cdot w_i\right]}{\sigma^3}
\end{equation}

\paragraphb{Spectral Kurtosis:}
Spectral Kurtosis gives a measure of the flatness or spikiness of a distribution relative to a normal distribution. It
is computed from the fourth central moment using the following function:\nolinebreak
\begin{equation}
Kurtosis = \frac{\sum_{i=1}^{N}\left[(f_i-\mu)^4 \cdot w_i\right]}{\sigma^4}
\end{equation}
A kurtosis value of 3 means the distribution is similar to a normal distribution whereas values less than 3 refer to flatter distributions and values
greater than 3 refers to steeper distributions.

\paragraphb{Spectral Flatness:}
Spectral flatness measures how energy is spread across the spectrum, giving a high value when energy is equally distributed and a low value when
energy is concentrated in a small number of narrow frequency bands. The spectral flatness is calculated by dividing the geometric mean of the power
spectrum by the arithmetic mean of the power spectrum~\cite{Johnston88}:\nolinebreak
\begin{equation}
Flatness=\frac {\left[\prod_{i=1}^N m_i\right]^{1/N}} {\frac{1}{N}\sum_{i=1}^N m_i}
\end{equation}
where $m_i$ represents the magnitude of bin number $i$. One advantage of using spectral flatness is that it is not affected by the amplitude of the 
signal.

\paragraphb{Spectral Brightness:}
Spectral brightness calculates the amount of spectral energy corresponding to frequencies higher than a given cut-off threshold. Spectral brightness 
can be computed using the following equation:
\begin{equation}
Brightness_{f_c} = \sum_{i=f_c}^{N}m_i
\end{equation}
where $f_c$ is the cut-off frequency (set to 1500Hz) and $m_i$ is the magnitude of the $i$-th frequency component of the
spectrum.

\paragraphb{Spectral  Rolloff:}
The spectral rolloff is defined as the frequency below which 85\% of the distribution magnitude is concentrated~\cite{Tzanetakis2002}
\begin{equation}
\argmin_{f_c\in\{1,\dots,N\}}\sum_{i=1}^{f_c}m_i \geq 0.85\cdot\sum_{i=1}^{N}m_i
\end{equation}
where $f_c$ is the rolloff frequency and $m_i$ is the magnitude of the $i$-th frequency component of the spectrum. 

\paragraphb{Spectral Irregularity:}
Spectral irregularity measures the degree of variation of the successive peaks of a spectrum. This feature provides the ability to capture the jitter
or noise in spectrum. Spectral irregularity is computed as the sum of the square of the difference in amplitude between adjoining spectral
peaks~\cite{jensen1999timbre} using the following equation:\nolinebreak
\begin{equation}
Irregularity = \frac{\sum_{i=1}^N (a_i - a_{i +1})^2 }{\sum_{i=1}^N a_i^2 }
\end{equation}
where the $(N+1)$-th peak is assumed to be zero. A change in irregularity changes the perceived timbre of a sound.

\paragraphb{Spectral Flux:}
Spectral flux is a measure of how quickly the power spectrum of a signal changes. It is calculated by taking the average Euclidean distance between 
the power spectrum of two contiguous frames.

\paragraphb{Spectral Attack Time:}
This features computes the average rise time to spectral attacks where spectral attacks are local maxima in the spectrum~\cite{mirtoolbox-manual}. 

\paragraphb{Spectral Attack Slope:}
This features computes the average slope to spectral attacks where spectral attacks are local maxima in the spectrum~\cite{mirtoolbox-manual}. 

\section{Screenshot of Our Data Collection Webpage}
\begin{figure}[!htb]
\centering
\begin{tabular}{cc}
\epsfig{file=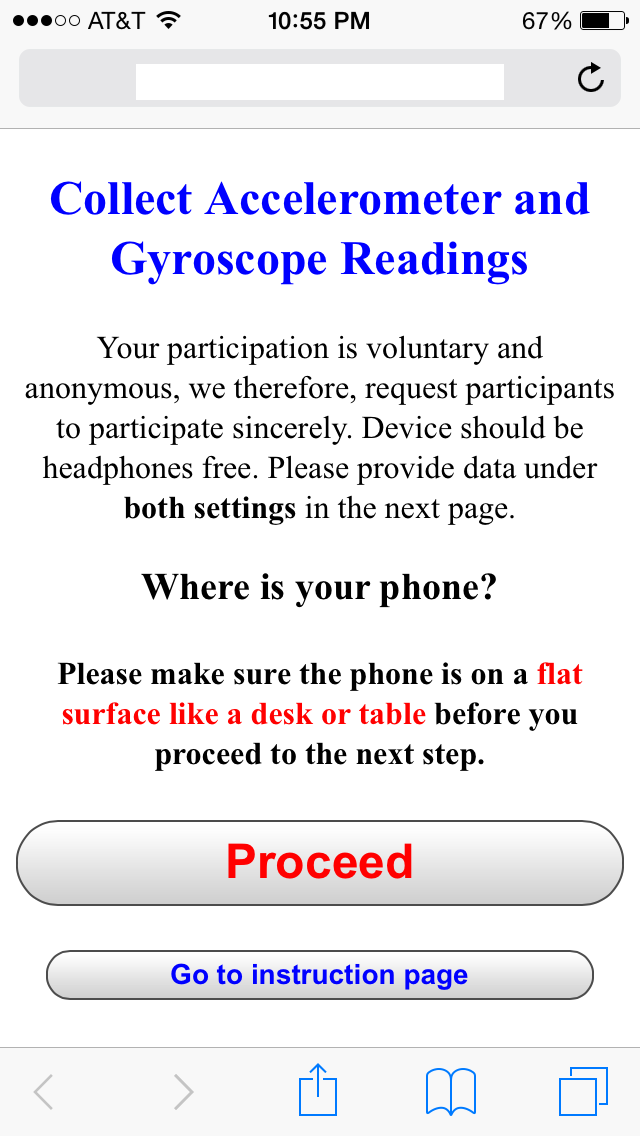,width=0.45\columnwidth,clip=}&\epsfig{file=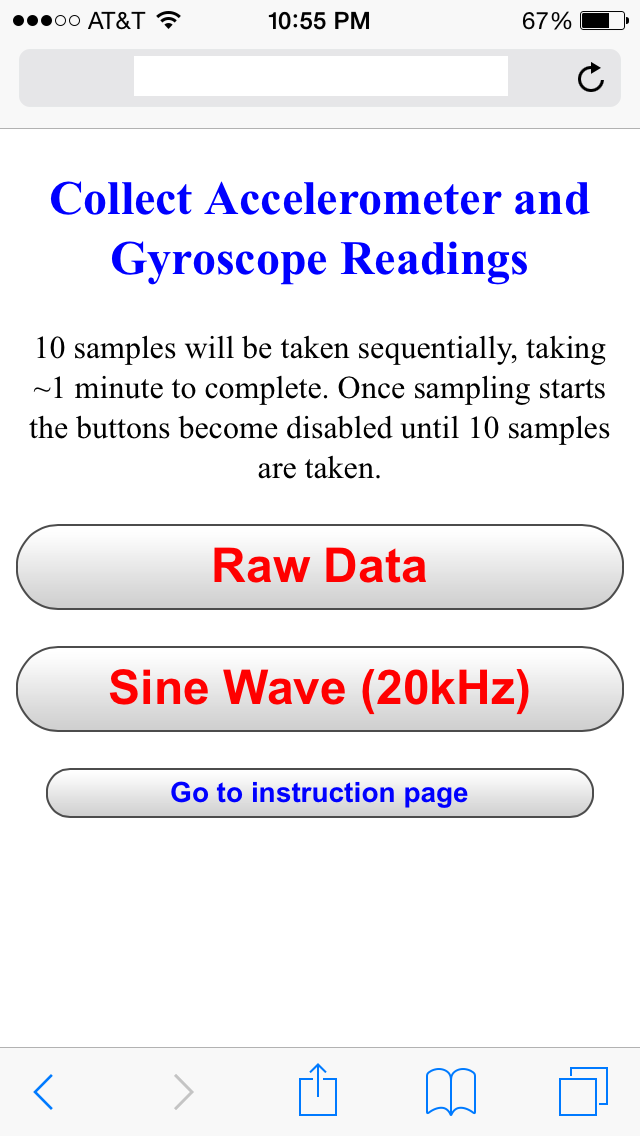,width=0.45\columnwidth,clip=}
\end{tabular}
\caption{Screenshot of our data collection website.} 
\label{website}
\end{figure}

{\label{appendix_screenshot}}
We provide screenshots (see figure~\ref{website}) of our data collection website to give a better idea of how 
participants were asked to participate. 

\section{Accessing Motion Sensors From Browser}{\label{appendix_code}}
To access motion sensors the \emph{DeviceMotion} class needs to be initialized. A sample JavaScript snippet is given below:
{\footnotesize
\begin{verbatim}
if(window.DeviceMotionEvent!=undefined){
    window.addEventListener('devicemotion', motionHandler);
    window.ondevicemotion = motionHandler;
}
function motionHandler(event) {     
    agx = event.accelerationIncludingGravity.x;
    agy = event.accelerationIncludingGravity.y;
    agz = event.accelerationIncludingGravity.z;             
    ai = event.interval;
    rR = event.rotationRate;
    if (rR != null) {
            arAlpha = rR.alpha;
            arBeta = rR.beta ;
            arGamma = rR.gamma;
    }
}
\end{verbatim}
}